%% file: AP_Paper2.tex
\documentclass[useAMS,a4paper,usenatbib]{mnras}

\usepackage{newtxtext}

\usepackage[T1]{fontenc}
\usepackage{ae,aecompl}

\usepackage{graphicx}	
\usepackage{float}
\usepackage{caption}

\usepackage{amsmath}	
\usepackage{amssymb}	
\usepackage{listings}
\usepackage{color}
\usepackage{booktabs}
\usepackage{natbib}

\title[Formation of Planetary Populations II]{Formation of Planetary Populations II: Effects of Initial Disk Size \& Radial Dust Drift}

\author[M. Alessi, R. Pudritz, \& A. Cridland]{
Matthew Alessi$^{1}$\thanks{E-mail:
alessimj@mcmaster.ca (MA); pudritz@mcmaster.ca (REP); cridland@strw.leidenuniv.nl (AJC)},
Ralph E. Pudritz$^{1,2}$\footnotemark[1], and Alex J. Cridland$^{3}$\footnotemark[1]\\
$^{1}$Department of Physics and Astronomy, McMaster University, Hamilton, ON L8S 4M1, Canada\\
$^{2}$Origins Institute, McMaster University, Hamilton, ON L8S 4M1, Canada\\
$^{3}$Leiden Observatory, Leiden University, 2300 RA Leiden, the Netherlands
}

\date{Accepted XXX. Received YYY; in original form ZZZ}

\pubyear{2020}

\begin{document}
\label{firstpage}
\pagerange{\pageref{firstpage}--\pageref{lastpage}}
\maketitle

\begin{abstract}

Recent ALMA observations indicate that while a range of disk sizes exist, typical disk radii are small, and that radial dust drift affects the distribution of solids in disks. Here we explore the consequences of these features in planet population synthesis models. A key feature of our model is planet traps - barriers to otherwise rapid type-I migration of forming planets - for which we include the ice line, heat transition, and outer edge of the dead zone. We find that the ice line plays a fundamental role in the formation of warm Jupiters. In particular, the ratio of super Earths to warm Jupiters formed at the ice line depend sensitively on the initial disk radius. Initial gas disk radii of $\sim$50 AU results in the largest super Earth populations, while both larger and smaller disk sizes result in the ice line producing more gas giants near 1 AU. This transition between typical planet class formed at the ice line at various disk radii confirms that planet formation is fundamentally linked to disk properties (in this case, disk size), and is a result that is only seen when dust evolution effects are included in our models. Additionally, we find that including radial dust drift results in the formation of more super Earths between 0.1 - 1 AU, having shorter orbital radii than those produced in models where dust evolution effects are not included. 

\end{abstract}




\input{Introduction.tex}

\input{Model.tex}

\input{Results.tex}

\input{Discussion_Conclusion.tex}

\section*{Acknowledgements}
The authors thank Tilman Birnstiel and Yasuhiro Hasegawa for insightful discussions regarding this work. We also thank the anonymous referee for their helpful comments improving the quality of this paper. M.A. acknowledges funding from the National Sciences and engineering Research Council (NSERC) through the Alexander Graham Bell CGS/PGS Doctoral Scholarship and from an Ontario graduate scholarship. R.E.P. is supported by an NSERC Discovery Grant. A.J.C acknowledges financial support from the European Union A-ERC grant 291141 CHEMPLAN, by the Netherlands Research School for Astronomy (NOVA), and by a Royal Netherlands Academy of Arts and Science (KNAW) professor prize. This work made use of Compute/Caclul Canada. This research has made use of the NASA Exoplanet Archive, which is operated by the California Institute of Technology, under contract with the National Aeronautics and Space Administration under the Exoplanet Exploration Program.



\bibliographystyle{mnras}
\bibliography{research}


\appendix

\input{AppendixA}
\input{AppendixB}

\input{AppendixC}
\bsp	
\label{lastpage}
\end{document}

%% file: Introduction.tex
\section{Introduction}

The current wealth of exoplanetary data provides crucial constraints on the potential outcomes of planet formation. The current sample of nearly 4000 confirmed exoplanets \citep{Borucki2011, Batalha2013, Burke2014, Rowe2014, Morton2016} is consistently increasing as the \emph{K2} mission \citep{Crossfield2016, Livingston2018, Livingston2018b} and \emph{TESS} \citep{Gandolfi2018, Huang2018, Vanderspek2019} continue to discover and confirm even more exoplanets. The distribution of planets on the mass-semimajor axis (hereafter M-a) diagram reveals an immense amount of information that can significantly constrain planet formation theories. For example, exoplanet populations can be discerned from the structure in the planet distribution on the M-a diagram, and the diagram can be divided into zones that broadly define these various planet populations \citep{ChiangLaughlin2013, HP13}. A key question that arises from this data is, how do planets populate these regions of the M-a diagram?

\begin{figure}
\includegraphics[width = 0.45\textwidth]{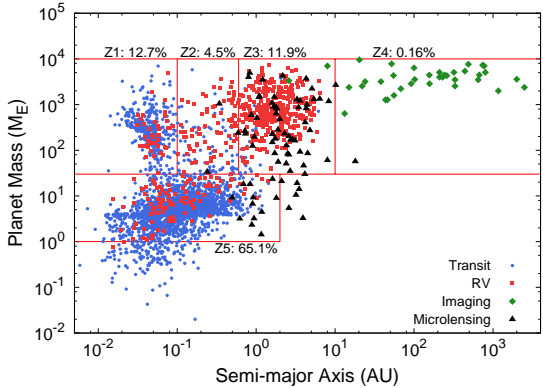}
\caption{The observed mass-semimajor axis distribution of confirmed exoplanets. As was first suggested by \citet{ChiangLaughlin2013}, the diagram is divided into zones that define planet populations: hot Jupiters (zone 1), period-valley giants (zone 2), warm Jupiters (zone 3), long-period giants (zone 4), and lastly super Earths and Neptunes (zone 5). The individual planets are colour-coded based on their initial detection technique. These data were compiled using the NASA Exoplanet Archive, current as of May 31, 2019.}
\label{Ma_Diagram}
\end{figure}

In figure \ref{Ma_Diagram}, we show the current distribution of confirmed exoplanets on the M-a digram. In terms of frequency, the dominant planet population consists of Earth-Neptune mass planets (1-30 M$_\oplus$) orbiting within 2 AU of their host stars, lying within zone 5 on the diagram (comprising 65.1\% of the total exoplanet population). Zones 1-4 define the various classes of gas giants: hot Jupiters (zone 1; 12.7\%), period-valley giants (zone 2; 4.5\%), warm Jupiters (zone 3; 11.9\%), and long-period giants (zone 4; 0.16\%). 

While the M-a diagram is useful in revealing the outcomes of planet formation, the distribution is shaped, in part, by the inherent biases present in exoplanet detection techniques. For example, the frequency of hot Jupiters on the M-a diagram exceeds the frequency of zone 3 planets, even though warm Jupiters have been shown to be the most common type of gas giant \citep{Cumming2008}. Occurrence rate studies, such as \citet{Santerne2016, Petigura2018}, account for detection biases to reveal the true underlying distribution of exoplanets. Results from these studies are extremely useful in constraining planet formation theories, offering the best means to compare theory to observations.

In this paper, we consider planet formation in the framework of the core accretion scenario \citep{Pollack1996}. Outcomes of the core accretion model have been shown to sensitively depend on host star and disk properties \citep{IdaLin2008, Mordasini2009, HP11, Alessi2017}.

We utilize the technique of planet population synthesis to link the range of outcomes of planet formation, as contained in the M-a diagram, to the variance of disk properties. Planet population synthesis allows one to explore the effects of ranges of input parameters on planet formation results, and the parameters we consider in this work are the disk's mass, lifetime, and metallicity. The technique has been used in many previous works, such as \citet{IdaLin2004, IdaLin2008, Mordasini2009, HP13, Bitsch2015, AliDib2017, Alessi2018}.  

Recent observations by \emph{ALMA} (e.g. \citet{ALMA2015, Andrews2018b}) and \emph{SPHERE} (e.g. \citet{Avenhaus2018}) have revolutionized our understanding of protoplanetary disks. In particular, observations that aim to measure disk masses and radii continue to inform planet formation models, as these quantities set the midplane densities throughout the disk, thereby affecting planet formation timescales. 

For example, \citet{Ansdell2016, Ansdell2018} measured dust and gas disk masses\footnote{Recent work has shown that dust masses may not be accurately estimated from sub-mm observations of disks due to optical depth or dust scattering effects \citep{Zhu2019}.} and radii of protoplanetary disks in Lupus with \emph{ALMA}. These observed disks surrounding a range of host-stellar masses have gas radii in the range of 63-500 AU at an age of $\sim$ 1-3 Myr. We emphasize, however, that these large, extended disks are the exception and are not indicative of typical disks that are much more compact, as indicated by numerous observations showing dust radii of $\lesssim 20-30$ AU \citep{Barenfeld2016, Barenfeld2017, Cox2017, Hendler2017,Tazzari2017, Cieza2019, Long2019}.

The compact dust-distribution resulting from radial drift models predicts that (sub-) mm emission from disks be more compact than measurements of CO isotopologues that trace the gas distribution \citep{Facchini2017, Trapman2019}. The \citet{Ansdell2018} survey of the Lupus disks found that gas disk radii were typically between 1.5 and 3 times larger than dust disk radii. This offset can be explained by the differences in optical depths of gas and dust without the need to consider effects of radial drift \citep{Facchini2017}. Radial drift, however, is indicated in cases where there is a more severe discrepancy between the gas and dust disk radii (i.e. \citet{Facchini2019}).

Quantifying disk properties has also been approached numerically in \citet{Bate2018}, who used radiative hydrodynamic calculations to compute distributions of disk masses and radii resulting from protostellar collapse. This work shows that initial disk radii significantly larger than $\sim$ 70 AU are uncommon. Constraints on distributions of disk properties, revealed either observationally or from simulations of disk formation, improve population synthesis models and the predicted outcomes of planet formation.


In \citet{Alessi2018}, we performed a suite of population synthesis calculations that assumed a constant disk dust-to-gas ratio of 1:100 while exploring the effects of planet envelope opacity and disk metallicity. In these calculations, we found that our models were unable to produce low-mass, short-period super Earths. While our models did produce many low-mass planets in the super-Earth - Neptune mass range, the majority of these had orbital radii exceeding 2 AU, situated outside of the observable limit for planets of these masses. This result was insensitive to model parameters.
 
Here, we incorporate a more realistic dust treatment by directly modelling the radial drift of solid dust particles throughout to the disk's evolution. We account for dust evolution through coagulation, fragmentation, and most importantly, radial drift. Radial drift of solids throughout the disk can drastically change the disk's dust density profile, depleting outer regions of large grains \citep*{Brauer2008, Birnstiel2010}. The resulting distribution of solids affects solid accretion rates onto planets, in turn affecting planet formation outcomes in this work. Including dust evolution, therefore, will have a particularly large affect on the formation of super-Earths and Neptunes (whose masses are dominated by solids), and their resulting period distribution.


A crucial feature of planet formation theories is a physical means to prevent the loss of forming planetary cores by rapid type-I migration \citep*{Alibert2004, IdaLin2008, Mordasini2009}. A solution to this ``type-I migration problem'' is planet traps - locations of zero net-torque on forming planetary cores that arise at inhomogeneities or transitions in disks \citep{Masset2006, HP11}. As these are locations of zero net-torque on planetary cores that would otherwise experience rapid inward migration, planet traps are the most likely locations of planet formation within the protoplanetary disk.

The planet traps that we consider in this model are the water ice line (the location of an opacity transition), the outer edge of the dead zone (a transition in disk turbulence), and the heat transition (separating an inner viscously heated region from an outer region heated through stellar radiation). While there are other disk features that can result in planet traps, such as the inner edge of the dead zone \citep{Masset2006}, the dust sublimation front \citep{Flock2019}, or other volatile ice lines (such as CO$_2$ - see \citet*{CPA2019}), the three we include are the traps in the main planet-forming region of the disk. Planet traps have been previously used in population synthesis calculations, such as \citet{MP2007, HP13, HP14, Hasegawa2016, Alessi2018}, who show that including this set of traps and their range of radii can result in the formation of the various observed exoplanet classes.
 
The goal of this work is to study the effects of dust evolution and radial drift on the resulting distribution of planets. By including dust evolution effects, we compare how important dust drift is to planet formation by comparing the period distribution of resulting super Earth and Neptune planets to our previous work \citep{Alessi2018} that assumed a constant 1:100 dust-to-gas ratio. Additionally, we will explore the link between disk properties and the statistical distribution of planets on the M-a diagram. In particular, since the dust distribution in disks will depend on their initial sizes, we will explore the effect of the characteristic radius of initial protoplanetary disks upon the resulting planet populations. 

We have discovered an intriguing result, namely, that the ratio of warm Jupiters and super Earths formed at the ice line trap is physically linked to the initial disk radius. Warm Jupiters are produced in excess of super Earths in both small and large disks, with the largest super Earth population formed at intermediate disk sizes of roughly 50 AU. This result is only encountered when dust evolution and radial drift effects are included in our models. Since the exoplanet data clearly indicates low-mass planets to be the dominant planet population, intermediate disk sizes (producing the largest number of super Earths) provide us with the best populations to compare with observations. This result is supported by MHD-simulations of disk formation during protostellar collapse that produce disks comparable to this size, depending on the mass-to-magnetic flux rato of the collapsing region \citep{Masson2016}. Additionally, we find our models are able to produce super Earths with small orbital radii ($\sim$ 0.03 AU) due to radial drift and the resulting delayed growth at the dead zone trap. This is a region of the M-a diagram that the constant dust-to-gas ratio models of our previous work, \citet{Alessi2018}, was unable to populate. 

The remainder of this paper is arranged as follows. In section 2, we give an overview of our model, first describing our calculation of the disk's physical conditions and its evolution in 2.1. We then outline the evolution and resulting distribution of  dust in 2.2. In 2.3, we describe our model of planet formation and migration - notably the trapped type-I migration phase. In 2.4 we outline our population synthesis method. Our planet population results are shown in section 3. In section 4, we discuss our results and summarize this work's key findings.

%% file: Model.tex
\section{Model}

This section summarizes the model used in this work, that combines models of the structure of an evolving protoplanetary disk, growth and radial drift of dust particles, the core accretion model of planet formation, and planet migration in a population synthesis calculation. We stochastically vary four parameters in our population synthesis approach, three of which describe properties of protoplanetary disks whose distributions are observationally constrained. The fourth parameter that we vary in our population synthesis framework is the only intrinsic model parameter stochastically varied.

For a detailed description of our disk, planet formation, and migration models, we refer the reader to \citet{Alessi2017, Alessi2018}. The dust evolution model, as a new inclusion to our calculations, is covered in detail in section 2.2.

\subsection{Disk Model}

We compute protoplanetary disk structure and evolution using the \citet{Chambers2009} model. We briefly mention the key assumptions of this model in this section, and refer the reader to Appendix A for a more complete description.

The \citet{Chambers2009} model is a 1+1D model that evolves with time due to viscous accretion and photoevaporation. While, generally, disk evolution takes place due to a combination of MRI-turbulence and MHD-driven disk winds, the \citet{Chambers2009} model inherently assumes the former. As a result of this assumption, disks will spread as they evolve according to,
\begin{equation} \frac{R}{R_0} = \left(\frac{\dot{M}}{\dot{M}_0}\right)^{-6/19} \;, \end{equation}
with $R$ being the disk's size, which depends on the disk's changing accretion rate $\dot{M}$ throughout its evolution. 

We emphasize that the \citet{Chambers2009} disk models the evolution of the total (gas $+$ dust) surface density, so the disk size $R$ best corresponds to a gas disk radius. Our fiducial setting for the initial disk radius $R_0$ is 50 AU. We highlight this feature of the disk model as the setting of the initial disk radius and its affect on planet populations will be explored in detail in this work. The fiducial setting for the initial disk mass is $M_0 = 0.1$ M$_\odot$. This is a stochastically-varied parameter in our population synthesis models - see equation \ref{LogNormal}.

The disk midplane is heated through a generalized viscous accretion (dominant in the inner region) and radiative heating from the host-star (dominant in the outer region). These two different heating mechanisms lead to different surface density and temperature power-law indices depending on the dominant heating mechanism at the radius in question. The heat transition, a planet trap in our model, separates these two regimes. 

We refer the reader to figure \ref{DiskPlot} for the disk's accretion rate evolution, and radial profiles of the surface density and midplane temperature throughout the disk's evolution.

\subsection{Dust Evolution}

The main addition to our model in this work is the inclusion of dust evolution, as the disk dust-to-gas ratio was assumed to be 1:100 in previous works (\citet{Alessi2017, Alessi2018}). We use the \citet*{Birnstiel2012} two-population dust model that accounts for dust evolution through coagulation, fragmentation, and radial drift. These effects are crucial for interpreting modern disk images (i.e. \citet{Birnstiel2018}). 

The \citet{Birnstiel2012} model is itself a simplified version of a full dust simulation over a distribution of grain sizes, as it only considers two grain sizes (a small, monomer grain size and a large grain near the upper limit of the grain size distribution),  yet is able to reproduce the full simulation results of \citet*{Birnstiel2010}. The two-population model is thus advantageous as our population synthesis calculations benefit from its reduced computational cost. We have modified the \citet{Birnstiel2012} dust model such that the gas evolves according to the \citet{Chambers2009} disk model (section 2.1). The initial global dust-to-gas ratio input into the \citet{Birnstiel2012} dust model scales with metallicity as,
\begin{equation} \frac{\Sigma_d}{\Sigma_g} = f_{\rm{dtg},0} 10^{[\rm{Fe}/\rm{H}]} \, , \label{DiskDTG} \end{equation}
where $f_{\rm{dtg},0} = 0.01$ is the often-assumed setting for Solar metallicity. In \citet{Alessi2018}, the dust-to-gas ratio was assumed to be radially and temporally constant, and dust evolution was not considered.


Fragmentation (i.e. \citet{Blum2000}) and radial drift (i.e. \citet{Weidenschilling1977}) are barriers to the maximum size that grains can grow. By equating the relative velocity of grains to their fragmentation velocity, $u_f$, \citet*{Birnstiel2009} show the maximum grain size in the \emph{fragmentation-limited} case to be,
\begin{equation} a_{\rm{frag}} = f_f \frac{2}{3\pi}\frac{\Sigma_g}{\rho_s\alpha_{\rm{turb}}}\frac{u_f^2}{c_s^2} \, ,\label{FragLimit} \end{equation}
where $f_f$ is an order-unity parameter, $\Sigma_g$ is the gas surface density, and $\rho_s$ is the volume-density of solids. Analytical models of grain size distributions in \citet*{Birnstiel2011} find that most of the mass in large grains is contained in sizes slightly below the maximum fragmentation-limited grain size. The fragmentation parameter $f_f$ in equation \ref{FragLimit} is used to correct this offset. By comparing to their detailed simulations \citep{Birnstiel2010}, a best-fit setting of $f_f = 0.37$ was found in \citet{Birnstiel2012}.

\begin{figure*}
\includegraphics[width = 0.45\textwidth]{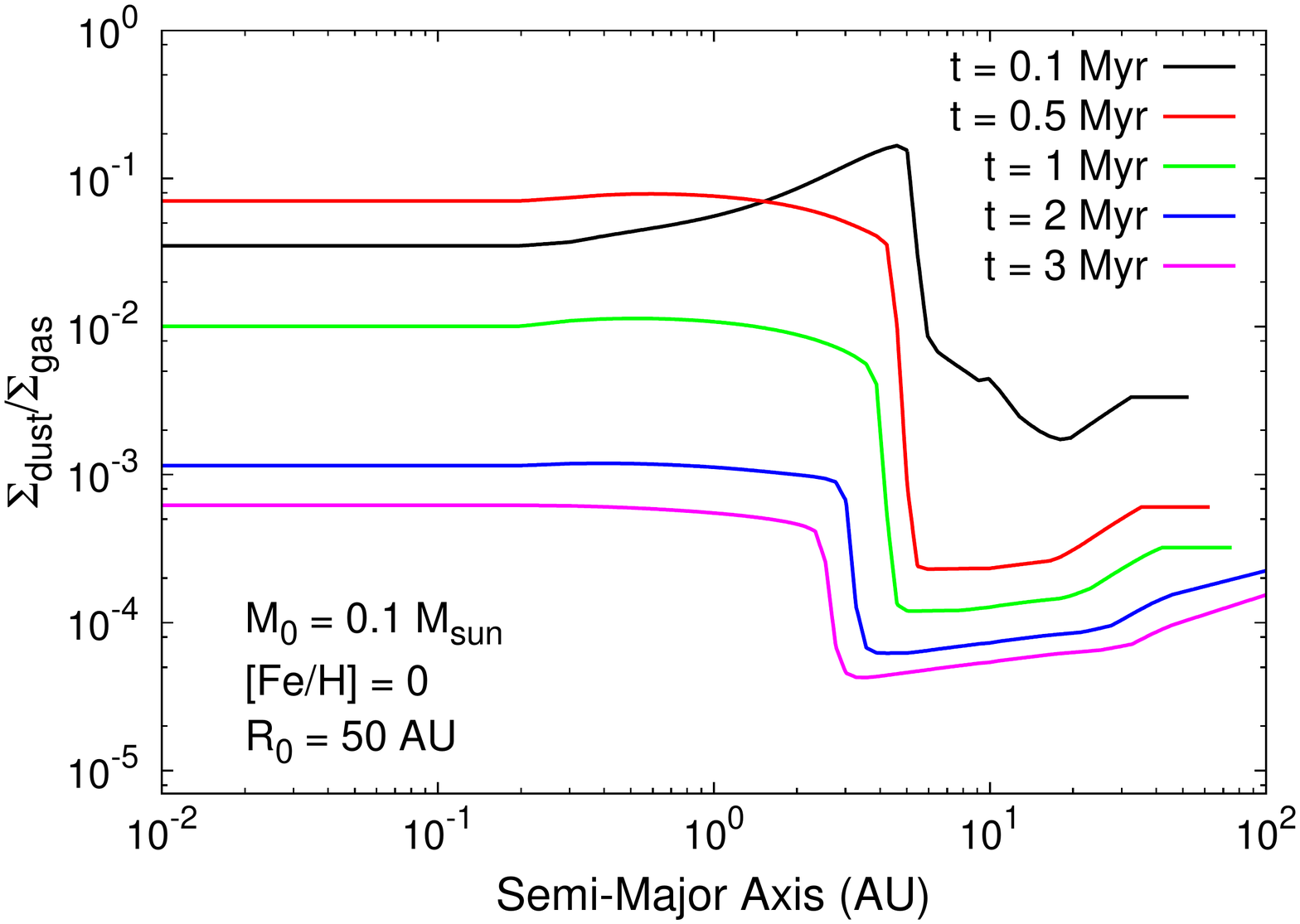} \includegraphics[width = 0.45\textwidth]{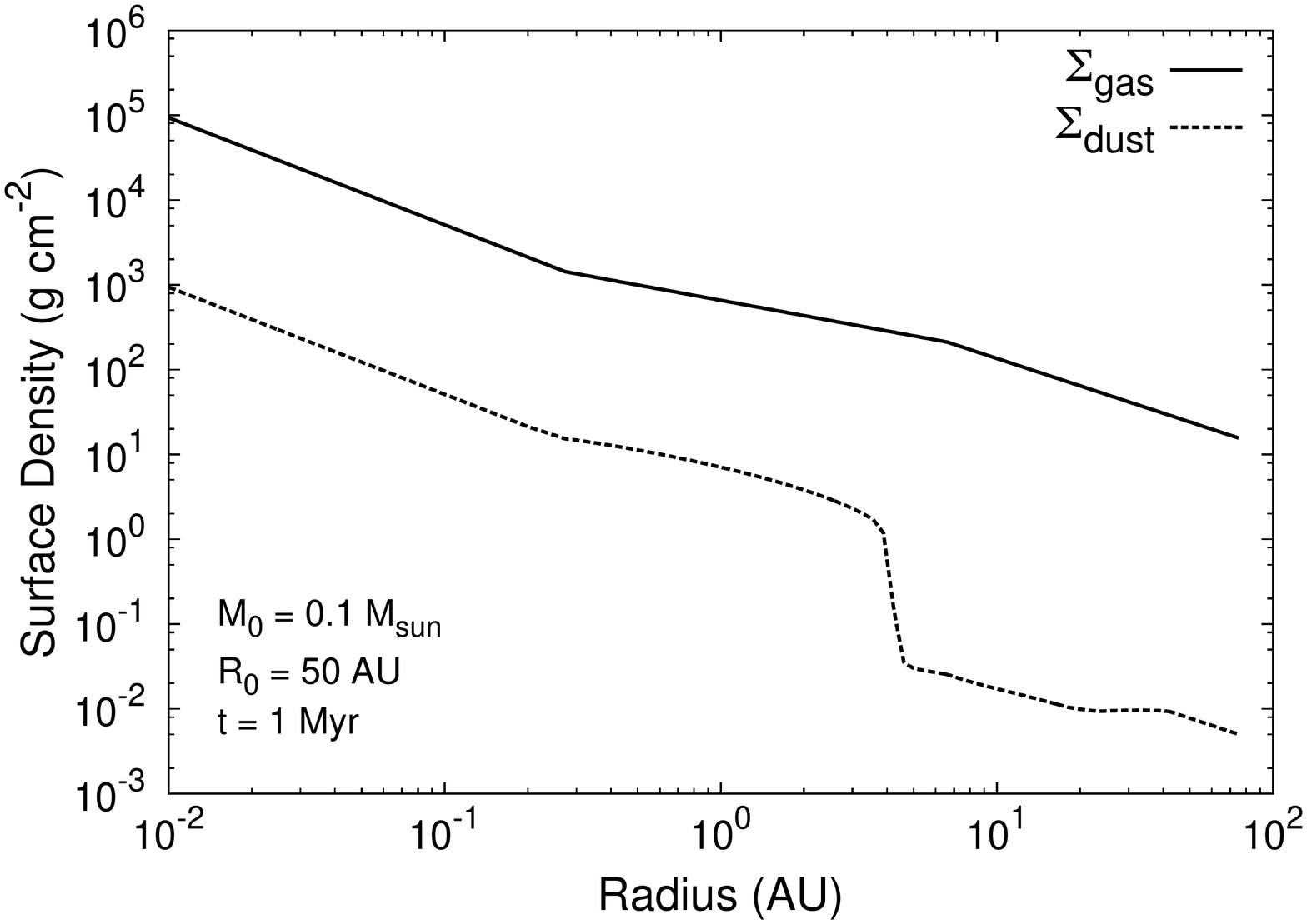}
\caption{\textbf{Left}: Dust-to-gas ratios computed using the \citet{Birnstiel2012} model are plotted for various stages our fiducial disk's evolution. The drift-limited region of the disk exterior to the snow line is apparent from these profiles. \textbf{Right}: The gas and dust surface densities at 1 Myr are shown.}
\label{DTG_Profiles}
\end{figure*}

We follow \citet{Birnstiel2010} for settings of the fragmentation velocity $u_f$. Within the ice line, grains have a fragmentation threshold velocity of 1 m s$^{-1}$, while outside the ice line, grains are enshrouded in an icy layer that strengthens the grains, increasing the fragmentation velocity threshold to 10 m s$^{-1}$. The region of the disk where water undergoes its phase transition spans of order a few tenths of an AU in our model \citep{Cridland2016, Alessi2017}. We follow \citet*{Cridland2017}, who model the transition in $u_f$ across the width of the ice line with an arctan function, fit to the radial ice distributions of \citet{Cridland2016}. 

Radial drift is a result of the drag forces experienced by dust grains due to the sub-Keplerian orbit of gas in the disk. While fragmentation does not change the radial distribution of dust surface density, $\Sigma_d$ (only redistributes dust mass among smaller grain sizes), radial drift affects the orbits of large grains which in turn affects $\Sigma_d$. The difference between the two effects is apparent when comparing dust evolution models, and resulting $\Sigma_d$ distributions, that include radial drift (i.e. \citet{Brauer2008}) to those that do not (i.e. \citet{Dullemond2005}).

By equating growth and radial drift timescales, \citet{Birnstiel2012} derive the maximum grain size in the \emph{drift-limited} case to be,
\begin{equation} a_{\rm{drift}} = f_d\frac{2\Sigma_d}{\pi\rho_s}\frac{V_K^2}{c_s^2} \gamma^{-1} \, , \label{DriftLimit} \end{equation}
where $f_d$ is an order-unity parameter, $V_K$ is the local Keplerian velocity, and $\gamma$ is the absolute value of the power-law index of the gas pressure profile,
\begin{equation} \gamma = \left | \frac{\rm{d}\, \ln P}{\rm{d}\, \ln r}\right| \,.\end{equation}
The parameter $f_d$ is calibrated in \citet{Birnstiel2012} by comparing to detailed numerical simulations \citep{Birnstiel2010}, who find a best-fit value of $f_d = 0.55$. We note that we have explored a range of settings $f_d = 0.1 - 1$ and find results of the \citet{Birnstiel2012} model to be insensitive to this parameter.

\begin{figure*}
\includegraphics[width = 0.32 \textwidth]{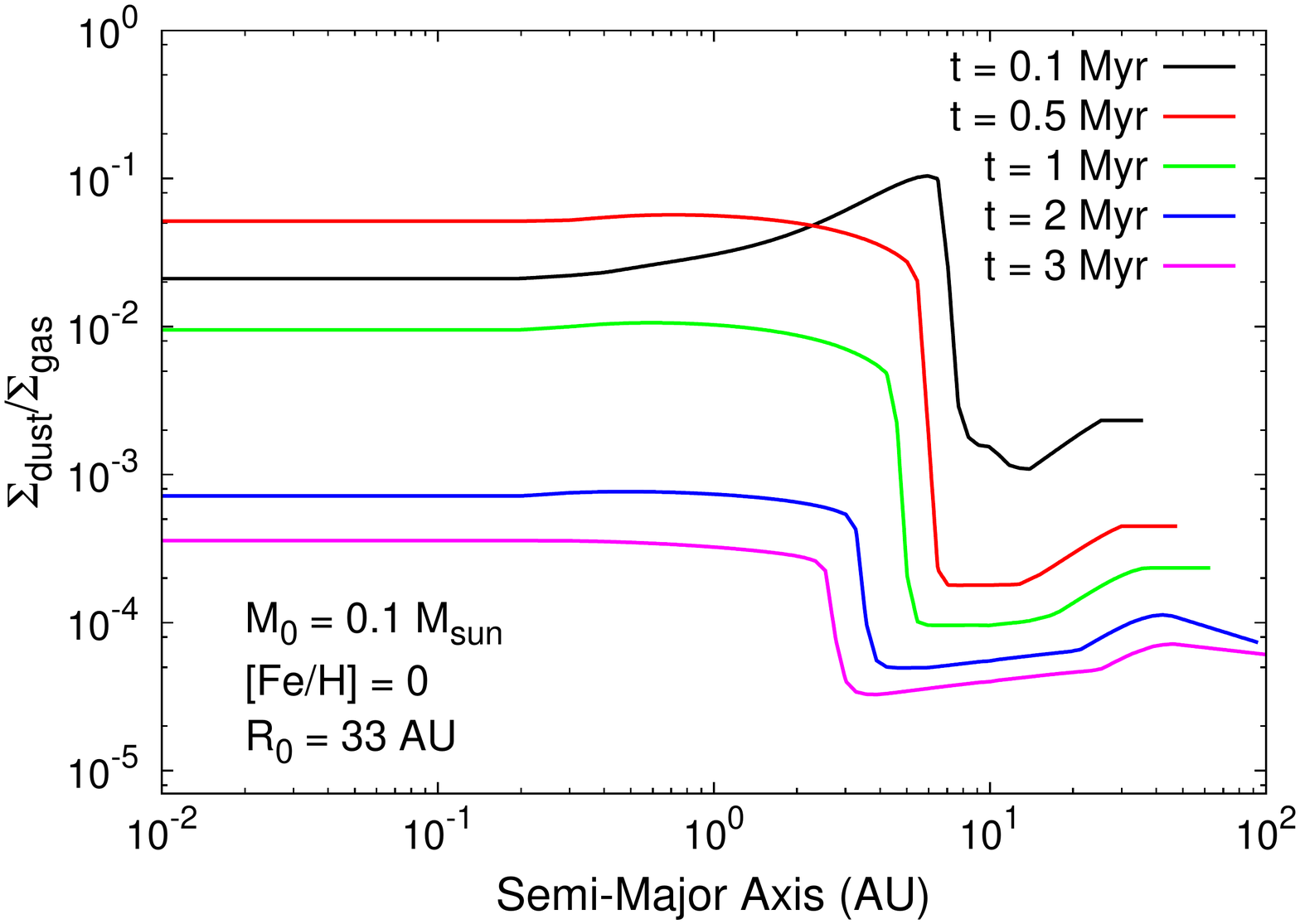} \includegraphics[width = 0.32 \textwidth]{Figures/DTG_50} \includegraphics[width = 0.32 \textwidth]{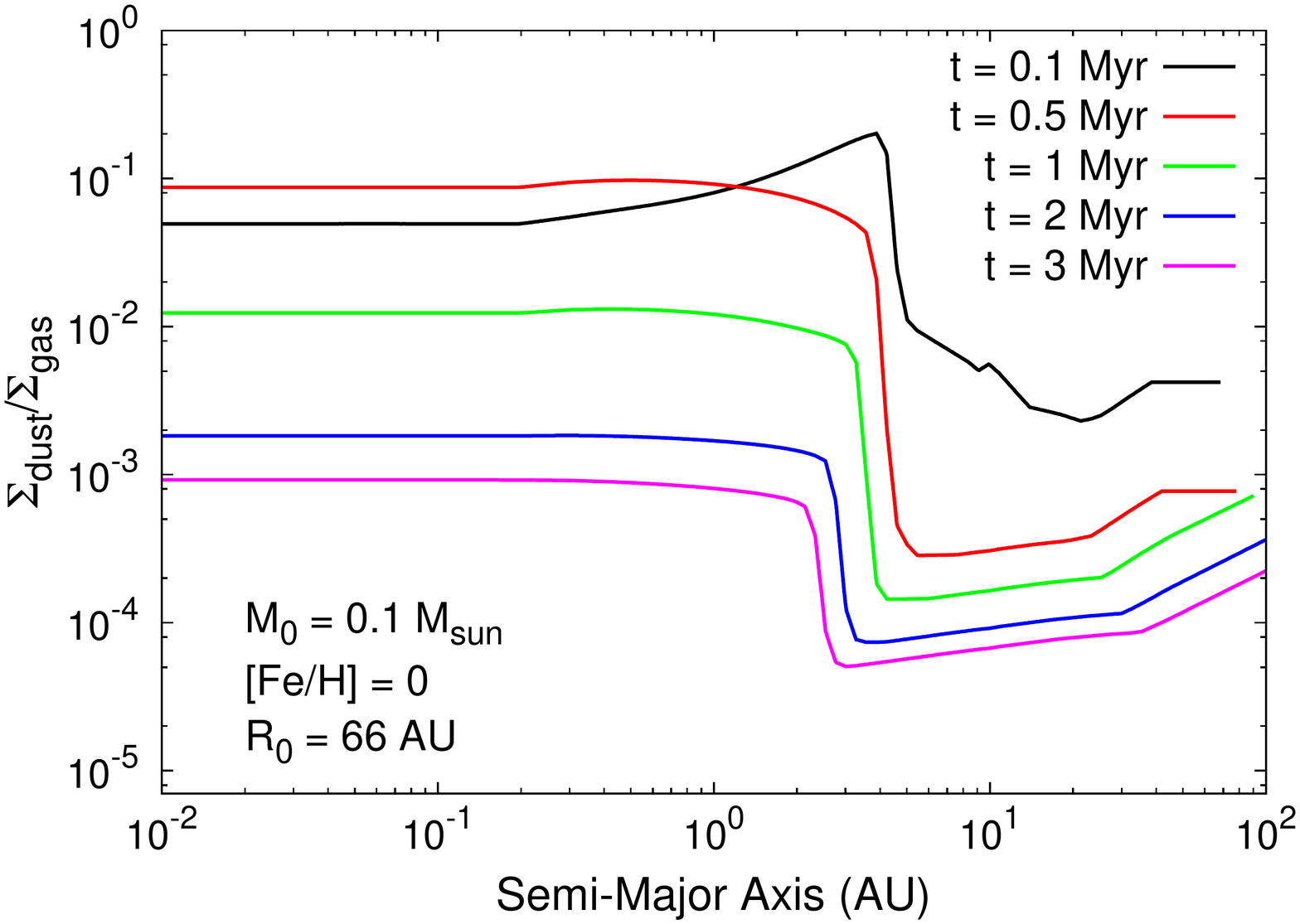}
\caption{Evolution of dust-to-gas ratio profiles are shown for disks of different initial radii: $R_0$ = 33 AU (left), the fiducial $R_0$ = 50 AU (middle), and $R_0$ = 66 AU (right). The dust-to-gas ratio in the inner regions (at the ice line and fragmentation-limited regime) is higher in disks with larger $R_0$ settings. Profiles otherwise show the same qualitative behaviour regardless of $R_0$ setting (fragmentation and drift-limited regimes with the ice line physically separating the two).}
\label{DTG_Size}
\end{figure*}

\begin{figure}
\includegraphics[width = 0.45 \textwidth]{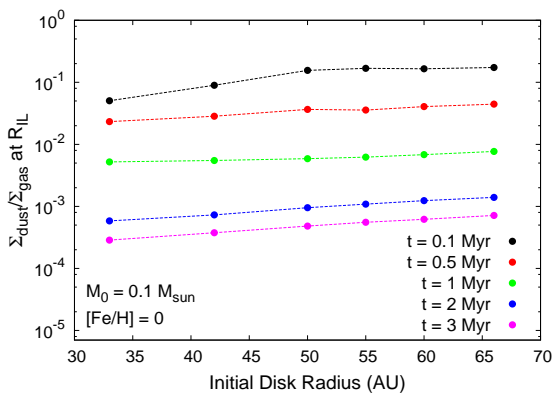}
\caption{The dust-to-gas ratio at the ice line is plotted at various times throughout fiducial mass and metallicity disks' evolutions with different settings of the initial disk radius. At all disk evolution times, the dust-to-gas ratio at the ice line is larger for disks with larger $R_0$ values.}
\label{DTG_Compare}
\end{figure}

The grain size distribution up to the maximum grain size can be reasonably fit with a power law (i.e. \citet{Birnstiel2011}),
\begin{equation} n(m)\,\rm{d}m = Am^{-\delta}\,\rm{d}m\,,\end{equation}
where $A$ and $\delta$ are positive constants. These constants depend on the maximum grain size, therefore depending on whether fragmentation or radial drift limits the growth (i.e. the smaller of $a_{\rm{frag}}$ and $a_{\rm{drift}}$). After computing the evolution of the two grain sizes in the two-population model, \citet{Birnstiel2012} reconstruct the full distribution, calibrated by \citet{Birnstiel2010}, throughout the disk. 

In figure \ref{DTG_Profiles} (left), we plot radial profiles of the dust-to-gas ratio, $\Sigma_d / \Sigma_g$, computed using the \citet{Birnstiel2012} dust model at various stages throughout our fiducial disk's evolution. The disk can be divided into three regions: (1) interior to the ice line, the grains have a lower fragmentation velocity, and their growth is \emph{fragmentation limited}; (2) outside the ice line, the grains' larger fragmentation velocity allows growth to larger sizes, and growth is therefore \emph{drift limited}; and lastly (3) the small region across the ice line where the fragmentation velocity transitions. In the right panel of figure \ref{DTG_Profiles}, we show the gas and dust surface density profiles after 1 Myr of disk evolution.

The effects of radial drift are apparent in figure \ref{DTG_Profiles}, as the regions of the disk outside the ice line (the drift-limited regime) are depleted in solids compared to within the ice line. At early times, the dust-to-gas ratio is enhanced near the ice line as radial drift efficiently moves solids from the outer disk inwards. The global dust-to-gas ratio decreases in time as stellar accretion takes place, and dust is removed without being replenished. After $\sim$ 1 Myr, the dust-to-gas ratio falls beneath the often-assumed 1:100 value at all radii, even in the fragmentation limited inner disk. 

From the results of the dust model, one can infer two imposed restrictions on our planet formation calculations, and in particular the solid accretion phase. First, and most crucially, solid accretion from regions of the disk outside the ice line will be inefficient as this region is depleted in solids by radial drift. The second is that solid accretion timescales will increase after $\sim$ 1 Myr as the dust-to-gas ratio has decreased beneath 0.01 across all radii. 

In figure \ref{DTG_Size}, we show how the initial disk radius affects dust-to-gas ratio profiles by comparing smaller (33 AU) and larger (66 AU) settings of the initial disk radius to the fiducial 50 AU case. Regardless of the initial disk radius setting, the computed dust-to-gas ratio profiles show the same qualitative behaviour. All profiles display a dust-depleted outer (drift-limited) region and an inner fragmentation-limited region with higher dust surface densities, with the ice line physically separating the two due to the changing fragmentation velocity.

In figure \ref{DTG_Compare}, we summarize a key trend seen in figure \ref{DTG_Size} by plotting the dust-to-gas ratios at the ice line at various evolution times in disks with different initial radii. We see that the dust-to-gas ratios in the inner regions of the disk (namely at the ice line and the fragmentation-limited region) are systematically higher in disks with larger initial disk radii settings. 

\begin{figure*}
\centering
\includegraphics[width = 0.45\textwidth]{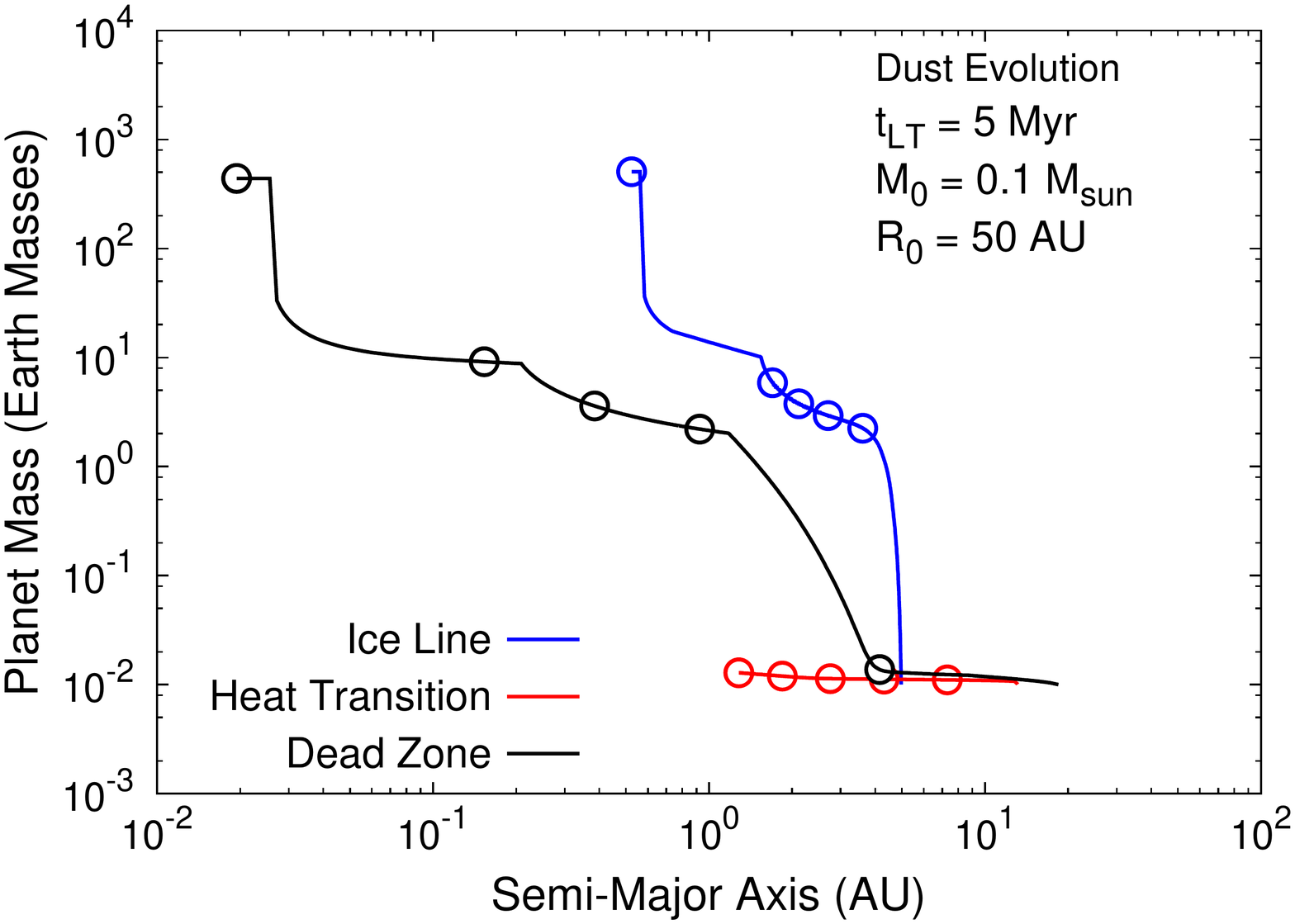} \includegraphics[width = 0.45\textwidth]{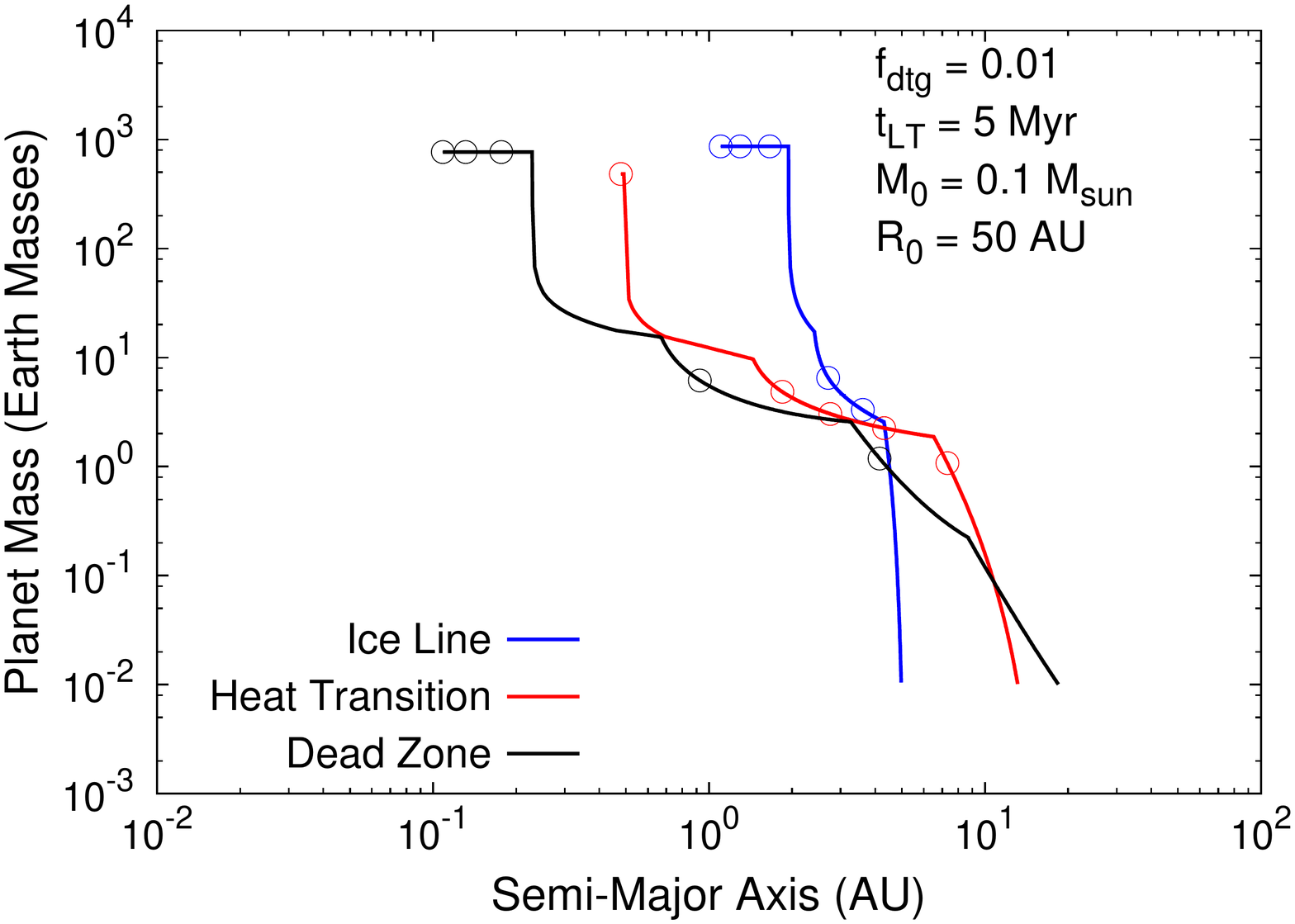}
\caption{\textbf{Left}: Planet formation tracks that include the effects of dust evolution on the solid surface density are shown for a long-lived, 5 Myr disk. Open circles along the tracks mark the location of the planets at 1 Myr intervals. The effects of the dust evolution model, namely low solid accretion rates outside the ice line, are apparent in the heat transition and dead zone tracks. \textbf{Right}: Planet formation tracks that assume a global, time-independent dust-to-gas ratio as was done in \citet{Alessi2018}, for comparison. The model set up is otherwise the same.}
\label{Tracks}
\end{figure*}

While the ice line does shift inwards slightly in larger disks due to the lower surface density, the change in the ice line's location between the three disk radius settings is small, with the difference being less than 1 AU between the 33 AU and 66 AU initial disk radii. Since, in the three models there is the same dust mass spread across the entire disk initially, larger disk radii settings will have more dust existing outside of the ice line simply because the disks themselves are more extended. The biggest effect of radial drift in these models is to remove dust from the outer disk, efficiently migrating it inwards to the ice line. Therefore, in more extended disks (bigger $R_0$), radial drift will have more material to transport inwards to the ice line, resulting in the trend of increasing dust-to-gas ratios in the inner disk with increasing initial disk radius.

A potential limitation of the \citet{Birnstiel2010, Birnstiel2012} models is that radial drift is too efficient, and the corresponding discrepancy between the dust and gas distributions in disks are too extreme. When comparing the spectral energy distribution indices resulting from the \citet{Birnstiel2010} simulation's dust distribution to observed indices of the Ophiucus disks, \citet*{Birnstiel2010b} found that radial drift needed to be suppressed in order to fit to the observations. Dust trapping by local pressure maxima in disks is a physical means by which radial drift can be halted and extended dust distributions be maintained \citep{Pinilla2012}. Recent disk observations have revealed dust substructures consistent with confinement to dust traps \citep{Casassus2015, vanderMarel2015, Dullemond2018}, supporting this theory.

Dust trapping is not included in this work, as it is not present in the \citet{Birnstiel2012} model. We do, however, include the effects of a changing fragmentation velocity across the ice line, which mimics the effects of a dust trap through local enhancement in solid density. While radial drift may be too efficient in this calculation, one of the main goals in this paper is to explore its unhindered effect on our planet populations. We also highlight that the combination of our previous work \citep{Alessi2018} where radial drift was not included, and this paper's high setting of radial drift explore the two extreme ends of radial drift's effects. 


\subsection{Planet Migration \& Formation}

Our treatment of planet migration and formation is unchanged from our previous work, \citet{Alessi2018}, and we refer the reader to Appendix B for a complete description.

The planet traps we include in our model are the water ice line, the heat transition, and the outer edge of the dead zone. The ice line's location is determined using an equilibrium chemistry calculation. The heat transition separates the inner portion of the disk where heating at the midplane takes place due to a generalized viscous accretion, and the outer portion of the disk heated via host-star radiation. The heat transition's location is determined within the framework of the \citet{Chambers2009} disk model. We compute the location of the dead zone's outer edge using the radiative transfer model presented in \citet{MP2003}. In this work, we only consider X-ray ionization caused by magnetospheric accretion and the resulting dead zone location, as our previous work (\citet{Alessi2018}) showed X-ray ionized disks (as opposed to galactic cosmic rays) to produce features in the resulting M-a distribution that better resembled the data.

\begin{table*}
\begin{centering}
\caption{Summary of model parameters}
\begin{tabular}{|c|c|c|}
\hline
\hline
Symbol & Meaning & Fiducial Value \\
\hline
&Population Synthesis Parameters& \\
\hline
$t_{\rm{LT}}$ &Disk lifetime& 3 Myr$^a$ \\
$M_0$ &Initial disk mass & 0.1 M$_\odot$$^b$\\
$[\rm{Fe}/\rm{H}]$ & Disk metallicity & 0$^c$\\
$f_{\rm{max}}$ & Maximum planet mass parameter (equation \ref{MaximumMass}) & 50$^d$ \\
\hline
&Disk Parameters& \\
\hline
$R_0$ & Initial gas disk radius & 50 AU \\
$\alpha$ & Effective viscosity coefficient (equations \ref{SS_Viscosity} \& \ref{EffectiveAlpha}) & 0.01 \\
$\tau_{\rm{int}}$ & Initial time (equation \ref{AccretionRate}) & 10$^5$ years \\
\hline
&Stellar Parameters$^e$& \\
\hline
$M_*$ & Stellar mass & 1 M$_\odot$ \\
$R_*$ & Stellar radius & 3 R$_\odot$ \\
$T_*$ & Stellar effective temperature & 4200 K \\
\hline
&Dust Model Parameters& \\
\hline
$f_{\rm{dtg},0}$ & Initial global dust-to-gas ratio at $[\rm{Fe}/\rm{H}]=0$ (equation \ref{DiskDTG}) & 0.01 \\
$f_f$ & Fragmentation parameter (equation \ref{FragLimit}) & 0.37$^f$ \\
$f_d$ & Drift parameter (equation \ref{DriftLimit}) & 0.55$^f$ \\
\hline
& Planet Formation Parameters$^g$ & \\
\hline
$f_{\rm{c,crit}}$ & Critical core mass parameter (equation \ref{CoreCrit}) & 1.26 \\
$c$ & Kelvin-Helmholtz $c$ parameter (equation \ref{Gas_Accretion}) & 7.7 \\
$d$ & Kelvin-Helmholtz $d$ parameter (equation \ref{Gas_Accretion}) & 2 \\
\hline
\hline
\end{tabular}
\label{ParameterTable}
\\
\begin{tabbing}
\textbf{Notes:}  \=\emph{a}. Log-normal distribution (equation \ref{LogNormal}) with $\mu_{\rm{lt}}$ = 3 Myr and $\sigma_{\rm{lt}}$ = 0.222. \\
\>\emph{b}. Log-normal distribution (equation \ref{LogNormal} with $\mu_{\rm{m}} = 0.1$ M$_\odot$ and $\sigma_{\rm{m}} = 0.138$. \\
\>\emph{c}. Normal distribution (equation \ref{Normal}) with $\mu_{\rm{Z}}$ = -0.012 and $\sigma_{\rm{Z}}$ = 0.21. \\
\>\emph{d}. Log-uniform distribution ranging from 1-500. \\
\>\emph{e}. Chosen to model a pre-main sequence solar type star \citep{Siess2000}.\\
\>\emph{f}. Parameters of \citet{Birnstiel2012} two-population dust model calibrated by fitting to full simulation of \citet{Birnstiel2010}.\\
\>\emph{g}. Determined using best-fit envelope opacity from \citet{Alessi2018}. 
\end{tabbing}
\end{centering}
\end{table*}


We consider the core accretion model of planet formation. We use the \citet{Birnstiel2012} dust model to compute the solid surface density distribution throughout the disk, thereby influencing solid accretion rates onto planetary cores. We use our best-fit envelope opacity models of \citet{Alessi2018} to set gas accretion parameters in equations for the critical core mass (equation \ref{CoreCrit}) and the Kelvin-Helmholtz timescale (equation \ref{Gas_Accretion}). Termination of gas accretion is handled by a parameter, $f_{\rm{max}}$, in our models (equation \ref{MaximumMass}) that relates a planet's final mass to its gap-opening mass (equation \ref{GapOpening}).




In figure \ref{Tracks}, left panel, we plot planet formation tracks resulting from a 5 Myr-lived disk that incorporate the dust model's effects on the solid distribution. We choose a long-lived disk to illustrate types of gas giants arising from various traps in our model. In figure \ref{Tracks}, right panel, we show planet formation tracks that assume a constant dust-to-gas ratio (the approach of \citet{Alessi2018}) for comparison. 

Planet formation at the ice line benefits from the early enhancement of solids caused by radial drift in the outer disk. This planet completes its solid accretion phase within 1 Myr and formation in this trap results in a warm Jupiter. The effects of radial drift on our planet formation model are apparent in the case of the heat transition track. The solid accretion rates onto this planet are extremely low, due to the trap being outside the ice line (see figure \ref{Traps}), in the radial-drift limited region of the disk that is depleted in solids.

The dead zone trap is initially situated outside the ice line as well, until a $\sim$ 1 Myr when it migrates within the ice line. Thus, the solid accretion rate is initially low for the planet forming in the dead zone trap as it is accreting from the radial-drift depleted region. After 1 Myr, the planet enters the fragmentation-limited region interior to the ice line with higher solid surface densities, and its accretion rate therefore increases. The result of planet formation in this trap is a hot Jupiter, whose mass is somewhat below that of the warm Jupiter near 1 AU

We emphasize that in the ice line and dead zone planet formation tracks, the slow gas accretion phase takes 2-3 Myr which is comparable to a typical disk lifetime. This highlights the way in which super Earths and Neptunes form in our model - these are planets whose disks photoevaporate during their slow gas accretion phases. Namely, if we were to use an average disk lifetime of 3 Myr in our example calculation (figure \ref{Tracks}), the ice line and dead zone would both have formed a Neptune mass planet at different orbital radii. The comparable slow gas accretion timescales to typical disk lifetimes suggest that this outcome should be common in our calculations.

\subsection{Population Synthesis}

We use a planet population synthesis approach to account for the spread in disk properties in the outcomes of planet formation models and distribution of computed populations on the M-a diagram. We stochastically vary four parameters in our population synthesis calculations, three of which are disk properties resulting from protostellar collapse that are external to our calculation - the disk lifetime, initial mass, and metallicity. The fourth stochastically varied parameter is the $f_{\rm{max}}$ parameter that determines the mass where gas accretion terminates (see equation \ref{MaximumMass} and related discussion). This is the only parameter intrinsic to our model that is varied. We use a log-uniform distribution for $f_{\rm{max}}$ ranging from 1 to 500.

We use the same disk mass, lifetime, and metallicity distributions as \citet{Alessi2018}. In the cases of disk mass and lifetime, we use a log-normal distribution,
\begin{equation} P(X|\mu_x,\sigma_x) \sim \exp \left(- \frac{(\log(X)-\log(\mu_x))^2}{2\sigma_x^2}\right) \, , \label{LogNormal} \end{equation}
with $\mu_{\rm{lt}} = 3$ Myr and $\sigma_{\rm{lt}}$ = 0.222 as mean and standard deviation for the disk lifetime distribution, and $\mu_{\rm{m}} = 0.1$ M$_\odot$ and $\sigma_{\rm{m}} = 0.138$ for the initial disk mass distribution. A normal distribution is used for disk metallicities,
\begin{equation} P(X|\mu_x,\sigma_x) \sim \exp \left(- \frac{(X-\mu_x)^2}{2\sigma_x^2}\right) \, , \label{Normal} \end{equation}
with $\mu_{\rm{Z}}$ = -0.012 and $\sigma_{\rm{Z}}$ = 0.21 providing a fit the metallicity distribution of G-type planet-hosting stars.

In this work, we explore the effects of initial disk radius on resulting planet populations. We do this by keeping the initial disk radius constant within each population, rather than choosing a distribution of disk radii to stochastically sample over. We do so to highlight the differences comparing populations with different initial disk radii. Including this as an additional stochastically varied parameter in the populations would `wash out' the parameter's effects. We note that we assume the distribution of disk masses to be unchanged regardless of the choice of initial disk radius in each individual population. While observations do indicate a correlation between disk masses and radii \citep{Tazzari2017, Tripathi2017, Ansdell2018}, we treat these as separate, uncorrelated parameters to isolate the effects of varying the initial disk radius on our resulting planet populations, independent of changes in the disk mass distribution. 

In the more massive disks considered in our population synthesis framework, and particularly with smaller settings of the disk's initial radius, disks in our model can be gravitationally unstable at early times ($\lesssim$ several 10$^5$ years), but only at large radii ($\gtrsim$ 25-30 AU). The region where planet formation takes place ($\lesssim$ 10 AU, outside of which solid accretion rates are negligible) lies well within the gravitationally stable region for all disks considered.

The most extreme case for gravitational instability that can be encountered in our populations is an initial disk mass of 0.2 M$_\odot$ and initial radius of 33 AU, for which the disk is initially stable out to 25 AU. The gravitationally unstable inner boundary shifts outwards as the disk evolves until the disk is entirely stable by 0.8 Myr. However, due to the log-normal distribution of initial disk masses (equation \ref{LogNormal}), sampling such a large disk mass as considered in this example is rare, and typical disk masses encountered in our populations will have gravitationally unstable regions confined to even larger radii and earlier times.

We include a summary of parameters used in our calculations and their fiducial settings in table \ref{ParameterTable}. Our population synthesis calculations consist of a Monte Carlo method whereby the four varied parameters' distributions are stochastically sampled before computing a planet formation track (as listed in table \ref{ParameterTable}, the disk lifetime, initial disk mass, metallicity, and maximum planet mass parameter $f_{\rm{max}}$). To compute a population, we iterate this process 1000 times in each trap, for a total of 3000 planets in each population.

%% file: Results.tex
\section{Results}

\subsection{Fiducial Population}

\begin{figure*}
\centering
\includegraphics[width = 0.45\textwidth]{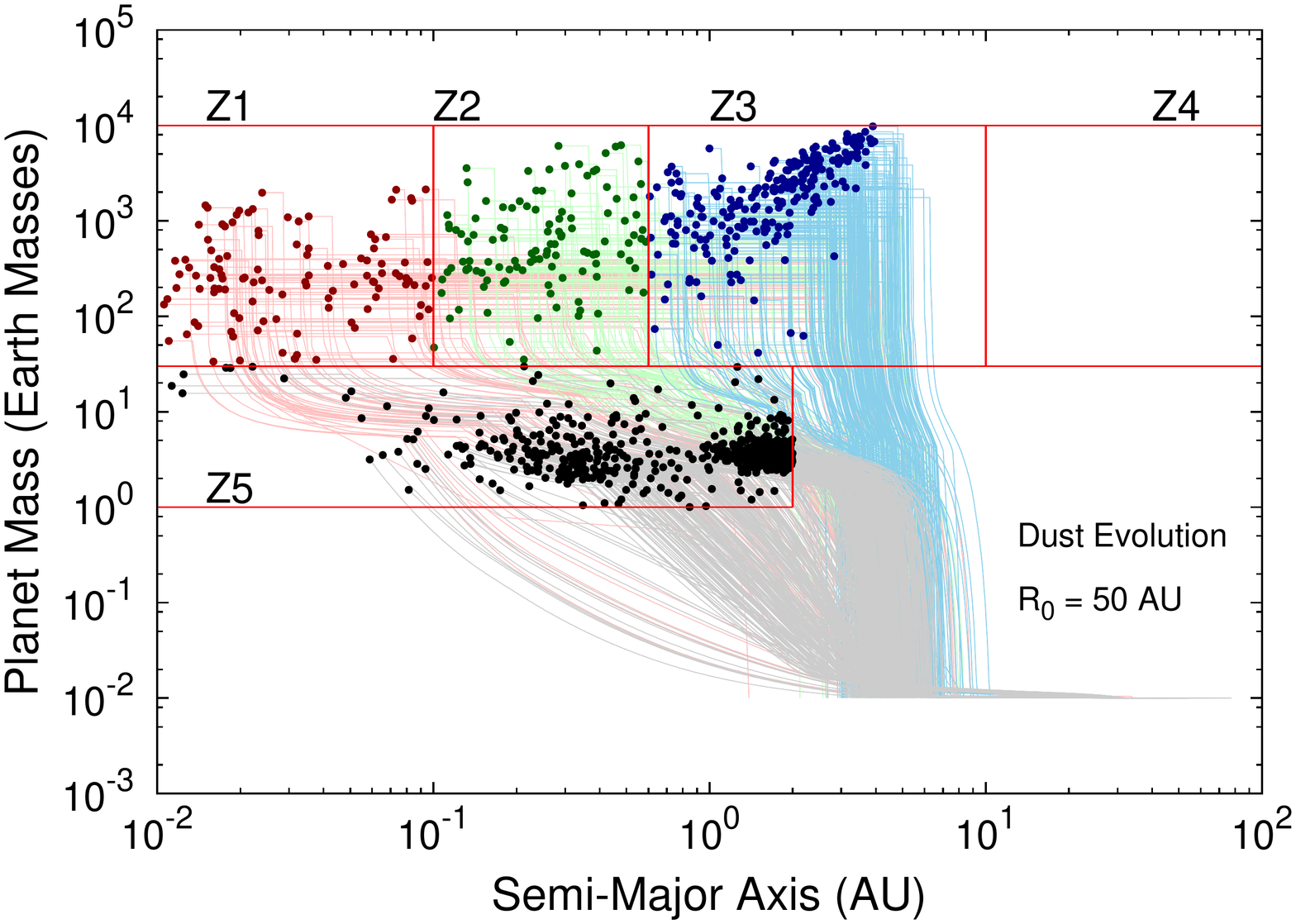} \includegraphics[width = 0.45\textwidth]{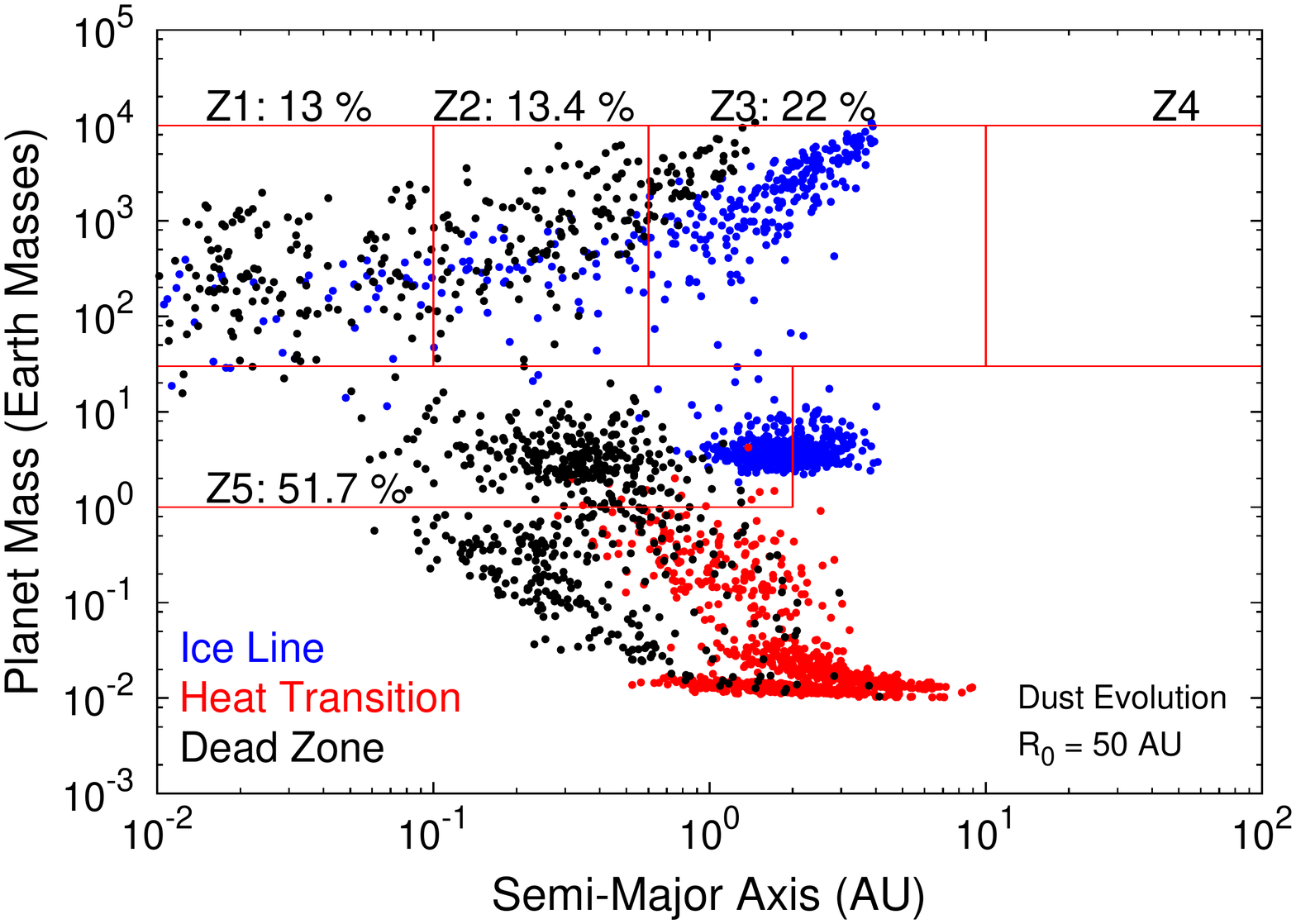}
\caption{The planet population resulting from the full dust evolution model and fiducial initial disk radius ($R_0 =$ 50 AU) is shown. \textbf{Left:} Planet formation tracks leading to the final population are shown only for planets that populate zones. End points of the tracks represent the final masses and semi-major axes of planets at the end of each of their disks' lifetimes. Colours of tracks and data points distinguish planets populating different zones of the diagram. \textbf{Right:} Resulting M-a distribution of the full population (including planets lying outside of the zones), with colour denoting the planet trap they formed in. We include the frequencies by which planets populate various zones.}
\label{50_Population}
\end{figure*}

In figure \ref{50_Population}, we show the population resulting from the full dust evolution treatment and the fiducial setting of the initial disk radius, $R_0$ = 50 AU. The data points show the final masses and semi-major axes of the planets at the disk lifetime of the disk in which they form - a varied parameter in our population synthesis calculation. The dust evolution model plays a key role in shaping this distribution, with the outer disk being depleted in solids by radial drift towards the ice line.  The resulting planet formation within each trap can be understood by considering where the traps exist with respect to the ice line.

Planet formation at the ice line in the fiducial model produces a mix of super Earths and Neptune-mass planets, as well as gas giants, primarily in the warm Jupiter (zone 3) region of the M-a diagram. At early stages in the disk's evolution, inward radial drift of solids from the outer disk results in a local enhancement of solids at the ice line (see figure \ref{DTG_Profiles}), and solid accretion at this trap is therefore efficient. Short solid accretion timescales in turn will result in short gas accretion timescales, making the ice line a main producer of warm gas giants.

In the case of planet formation at the heat transition, very few planets with masses exceeding only 1 M$_\oplus$ are formed. The majority of planets formed in this trap accrete very little mass, and have final planet masses near the initial condition mass of 0.01 M$_\oplus$. The heat transition lies outside the ice line for nearly all planet masses and metallicities encountered in the population synthesis calculations. Since the region of the disk outside of the ice line is depleted in solids by efficient radial drift, planets forming in the heat transition have extremely long solid accretion timescales due to the low solid surface densities, resulting in inefficient overall growth.

Planets formed in the dead zone trap result in a range of planet masses: low mass ($<$ 1 M$_\oplus$) planets, super Earths and Neptunes, as well as gas giants spread over a range of orbital radii, but typically shorter periods than gas giants formed in the ice line. The dead zone trap initially lies outside the ice line but quickly migrates inwards, intersecting the ice line at $\sim$ 1 Myr and ending up in the inner disk towards the end of disk evolution. Solid accretion is therefore inefficient initially while the dead zone exists outside the ice line, as was the case for planets forming in the heat transition. Solid accretion becomes efficient once the dead zone migrates within the ice line and planets forming within the dead zone encounter the high solid surface densities in the fragmentation-limited regime of the disk. 

The relatively fast migration of the dead zone trap and the delayed solid accretion caused by the dead zone's migration inside the ice line result in the three cases of planet classes: (1) Low-mass planets ($<$ 1 M$_\oplus$) are produced in the case of the shortest disk lifetimes where the planets primarily accrete from outside of the ice line where solids are depleted; (2) Zone 5 planets result from intermediate disk lifetimes where the dead zone has migrated inside the ice line and the solid accretion stage has taken place, but gas accretion has insufficient time to produce gas giants; Lastly (3), gas giants are formed from the dead zone in the longest-lived disks.

\subsection{Effects of Initial Disk Radius}

\begin{figure*}
\centering
\includegraphics[width = 0.97\textwidth]{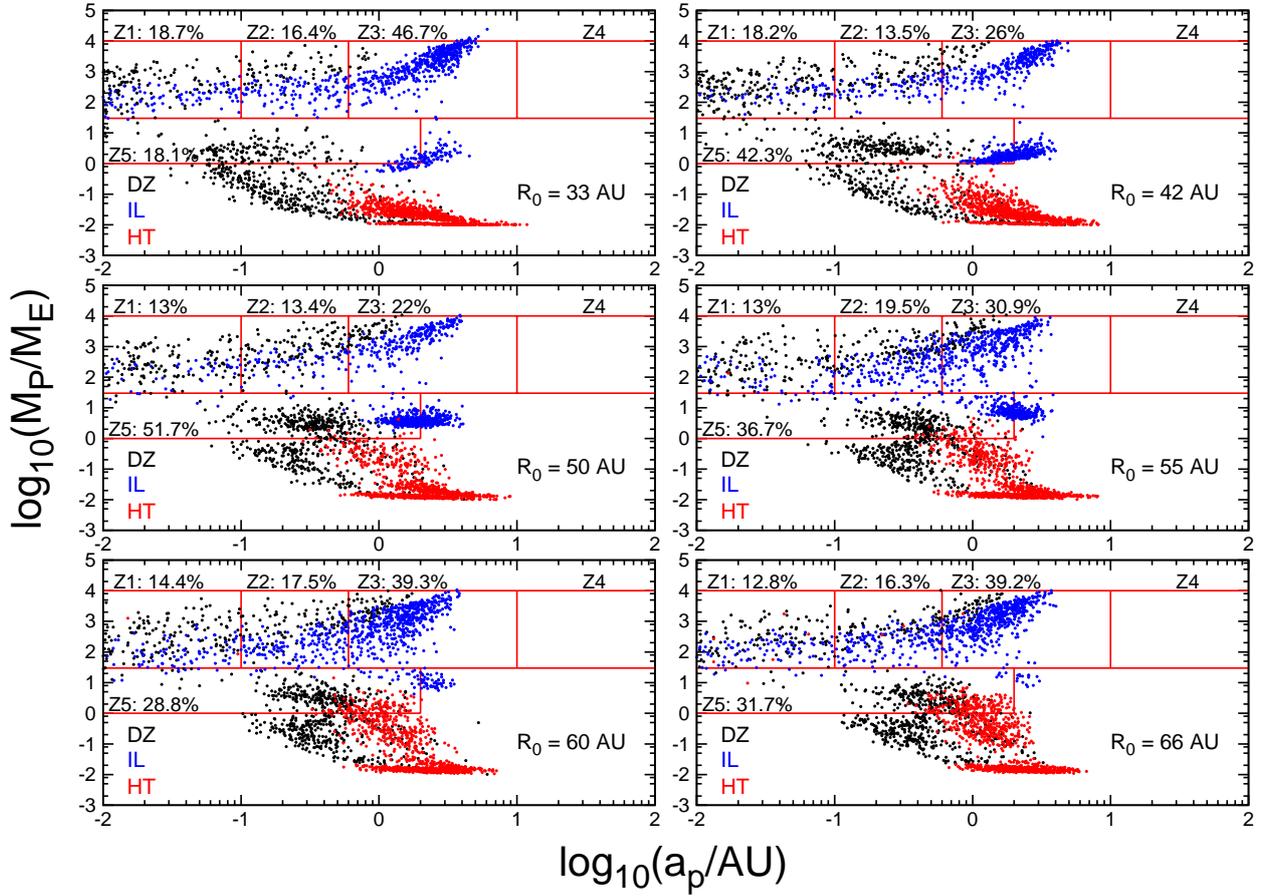}
\caption{M-a distributions of computed planet populations are shown for a range of initial disk radii, spanning from 33 AU to 66 AU. The largest super Earth and Neptune population is formed when considering intermediate disk sizes ($R_0$ = 50 AU), with smaller and larger initial disk sizes producing more warm gas giants.}
\label{Population_Radius}
\end{figure*}

\begin{figure}
\centering
\includegraphics[width = 0.45\textwidth]{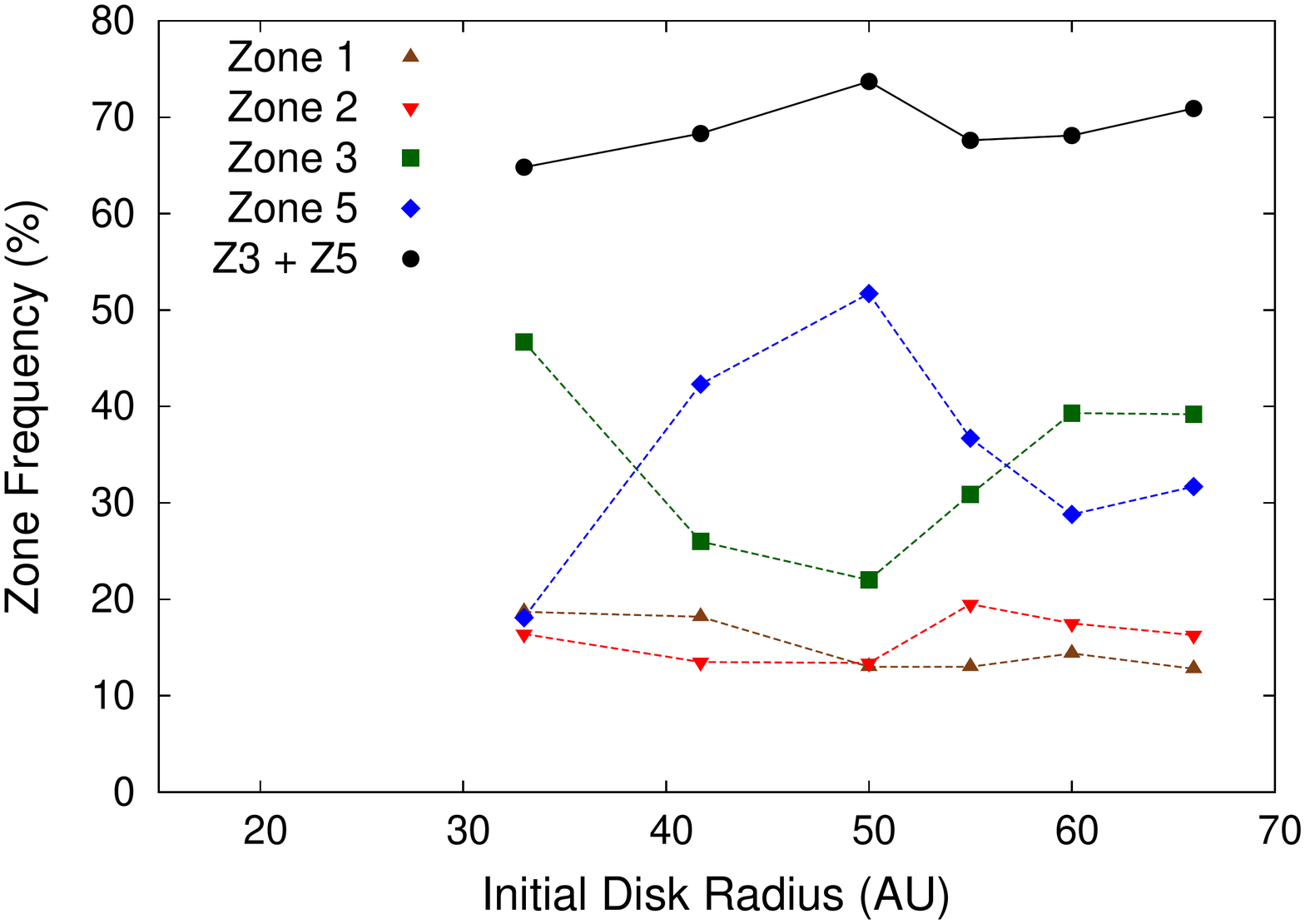}
\caption{We summarize the grid of computed populations in figure \ref{Population_Radius} by plotting the frequency by which planets populate zones as a function of initial disk radius. We also show a summed total of zones 3 \& 5 which remains relatively constant across the range of $R_0$ settings considered, showing that there is a trade-off between super Earths and warm Jupiter populations that each individually vary significantly as $R_0$ changes. The zone 1 and zone 2 populations show little variation with initial disk radius.}
\label{Zone_Frequencies}
\end{figure}

In figure \ref{Population_Radius}, we explore the effects of initial disk radius on our population results. We consider a range of initial disk radii spanning from 33 AU to 66 AU. This range was chosen to encompass the range of disk radii predicted by models of disk formation in protostellar collapse simulations, with the small disk radius end corresponding to a somewhat strong setting of the mass-to-magnetic flux parameter \citep{Masson2016}, and the large disk radius end corresponding to the pure hydrostatic case \citep{Bate2018}. Since our disk model is assumed to evolve via viscous evolution, through which disk spreading results. After 1 Myr of disk evolution, the corresponding range of disk sizes becomes 63 - 90 AU, and after 3 Myr of evolution (a typical disk lifetime), this range corresponds to 125-140 AU. As discussed in section 2.4, the range of disk masses that we consider remains the same in each population run, despite the initial disk radius changing. The changes in population outcomes between runs with different initial disk radii is therefore physically caused by changes in the disk's surface density.

The initial disk radius affects planet formation outcomes in each trap in our models, with the ice line being the most sensitive to the setting of $R_0$. The planets that form in the ice line trap can be divided into two groups: those in the super Earth - Neptune mass range, and gas giant planets, the majority of which populate the warm Jupiter region of the M-a diagram. In the cases of the smallest and largest disks considered in figure \ref{Population_Radius}, the ice line produces many more gas giants than zone 5 planets. The population of super Earths and Neptunes formed in the ice line reaches a maximum at intermediate disk radii near 50 AU. Additionally, the mass of the zone 5 planets that are produced in the ice line systematically increases as larger disk sizes are considered. In the case of the largest disk size, the population of ice line super Earths nearly washes out entirely, with the ice line producing gas giants almost exclusively.

The dead zone produces a combination of gas giants, zone 5 planets, and sub-Earth mass planets in each population regardless of the setting of initial disk radius. However, planet formation becomes slightly more efficient as the disk radius increases, for the same reason as it does in the heat transition. Thus, more gas giants are formed in the dead zone at larger $R_0$ settings. The minimum orbital radii of super Earths formed in the dead zone also increases with initial disk radius. In the smallest disk radius run, the dead zone produced super Earths with orbital radii as small as $\sim$ 0.03 AU, whereas in the case of the largest disk radius run, super Earths formed in the dead zone all had orbital radii larger than 0.1 AU.

The heat transition primarily produces sub-Earth mass planets in all but the largest initial disk radius runs. As larger disk sizes are considered, the upper end of the mass distribution of planets formed in the heat transition increases, and begins to substantially populate zone 5. In the $R_0$ = 60 AU and 66 AU runs, the heat transition forms a significant number of super Earths at larger orbital radii than those produced in the dead zone. Recalling that substantial solid accretion only takes place near or within the ice line, these results for the heat transition planets can be explained by noting that the heat transition trap converges with the ice line at systematically earlier times when lower surface density disks are considered. Therefore, it becomes increasingly likely for the heat transition to migrate to the high solid surface density regions of the disk at a given disk mass as the initial disk radius is increased. The subset of planets formed in the heat transition that incur some solid accretion thereby increase as the setting of $R_0$ increases.

In figure \ref{Zone_Frequencies}, we present the key plot of the paper, which summarizes the results of figure \ref{Population_Radius} by plotting the frequencies by which planets populate the different zones of the M-a diagram as a function of initial disk radius. 

We highlight the drastic variation among warm Jupiters (zone 3) and super Earths (zone 5) as the initial disk radius is changed. For both small and large settings of $R_0$, warm Jupiters form more frequently than super Earths, with super Earth formation frequency maximized at intermediate settings of initial disk radius near 50 AU. Moreover, there is a striking trade-off between these two planet populations, with the increasing super Earth population at intermediate disk radii coupled with a corresponding decreasing warm Jupiter population. 

We show this in figure \ref{Zone_Frequencies} by including a summed zone 3 and zone 5 population frequency that remains relatively constant across the range of explored $R_0$ settings. This disk radius-dependent exchange between super Earths and warm Jupiters is driven exclusively by planet formation at the ice line where the relative formation frequencies of super Earth and Neptune-massed planets and gas giants are sensitive to the setting of $R_0$. Our results show that planet formation is fundamentally linked to disk properties, as the formation frequency of super Earths is linked to the disk's radius. 

Additionally, we find that the populations of hot Jupiters (zone 1) and period-valley giants (zone 2) are insensitive to the setting of $R_0$, with the corresponding frequencies having minimal variation across the span of $R_0$ investigated. We find no disk radius-dependent interplay between hot Jupiters and super Earths comparable to that seen with the warm Jupiter population. Since the warm Jupiter population is to a large extent formed from the ice line and the hot Jupiters through the dead zone, we conclude that planet formation at the ice line is sensitive to the initial disk radius setting, while formation at the dead zone is not.

\begin{figure*}
\includegraphics[width = 0.45\textwidth]{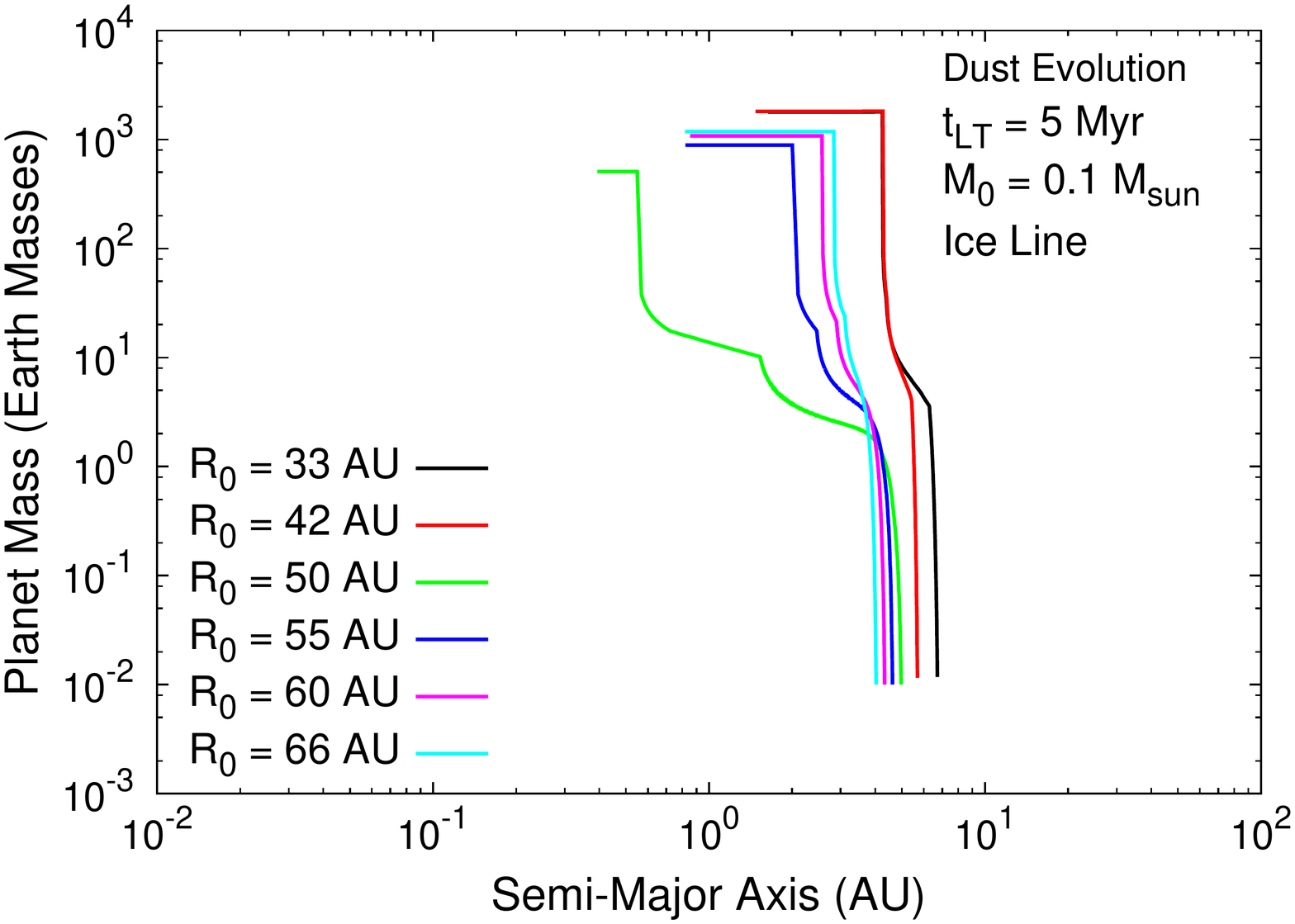} \includegraphics[width = 0.45\textwidth]{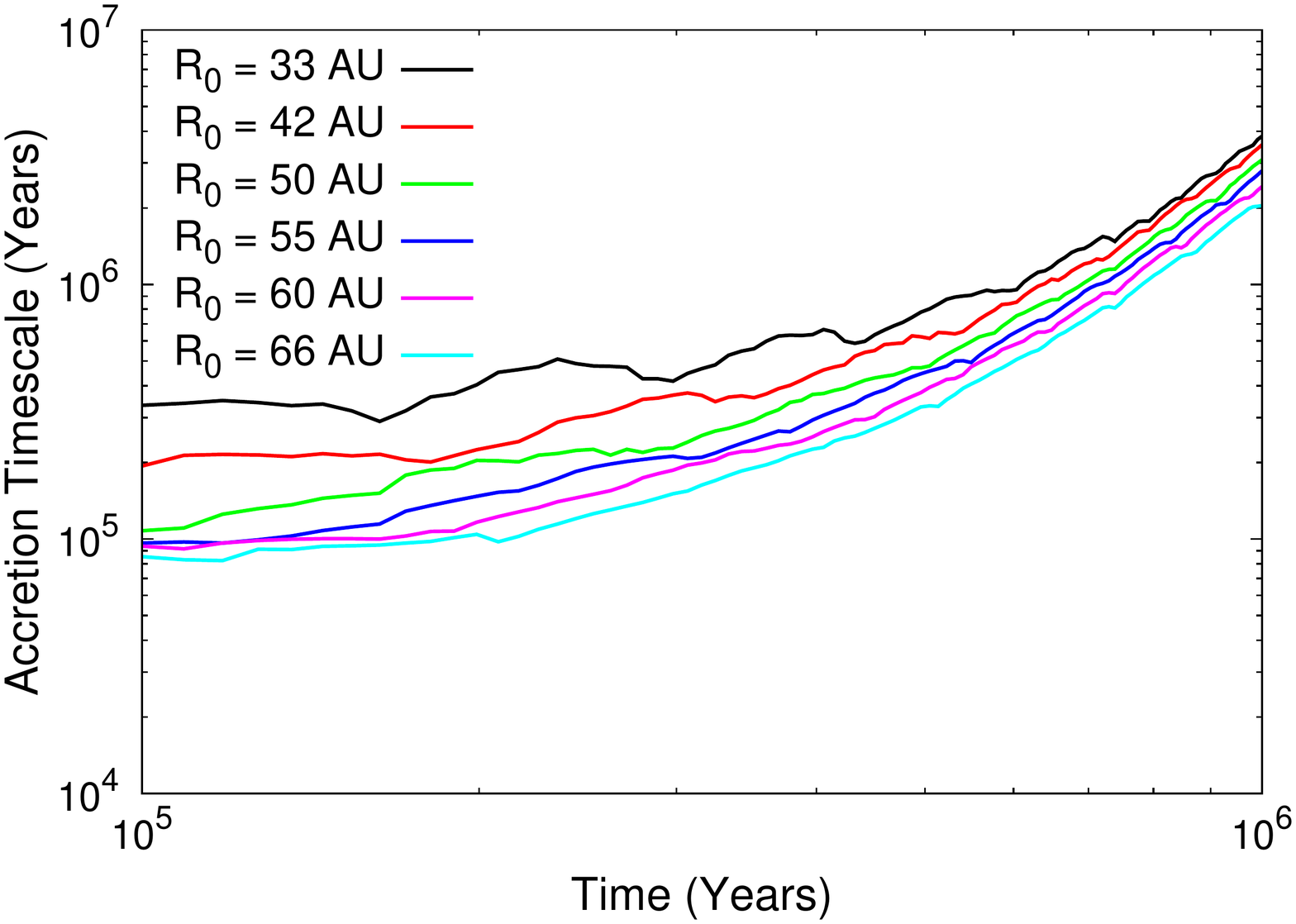}
\caption{\textbf{Left:} Planet formation tracks at the ice line are shown for a series of initial disk radii. The disk mass and metallicity are set at their fiducial values ($M_0$ = 0.1 M$_\odot$, [Fe/H] = 0). The initial position of the planetary cores (the position of the ice line) shifts slightly inwards for larger $R_0$ settings due to the lower column densities. \textbf{Right:} Solid accretion timescale, computed using equation \ref{Solid_Accretion} is plotted for the ice line planet formation tracks. The accretion timescales systematically decrease as the initial disk radius is increased.}
\label{Radius_Compare}
\end{figure*}

\begin{figure}
\includegraphics[width = 0.45 \textwidth]{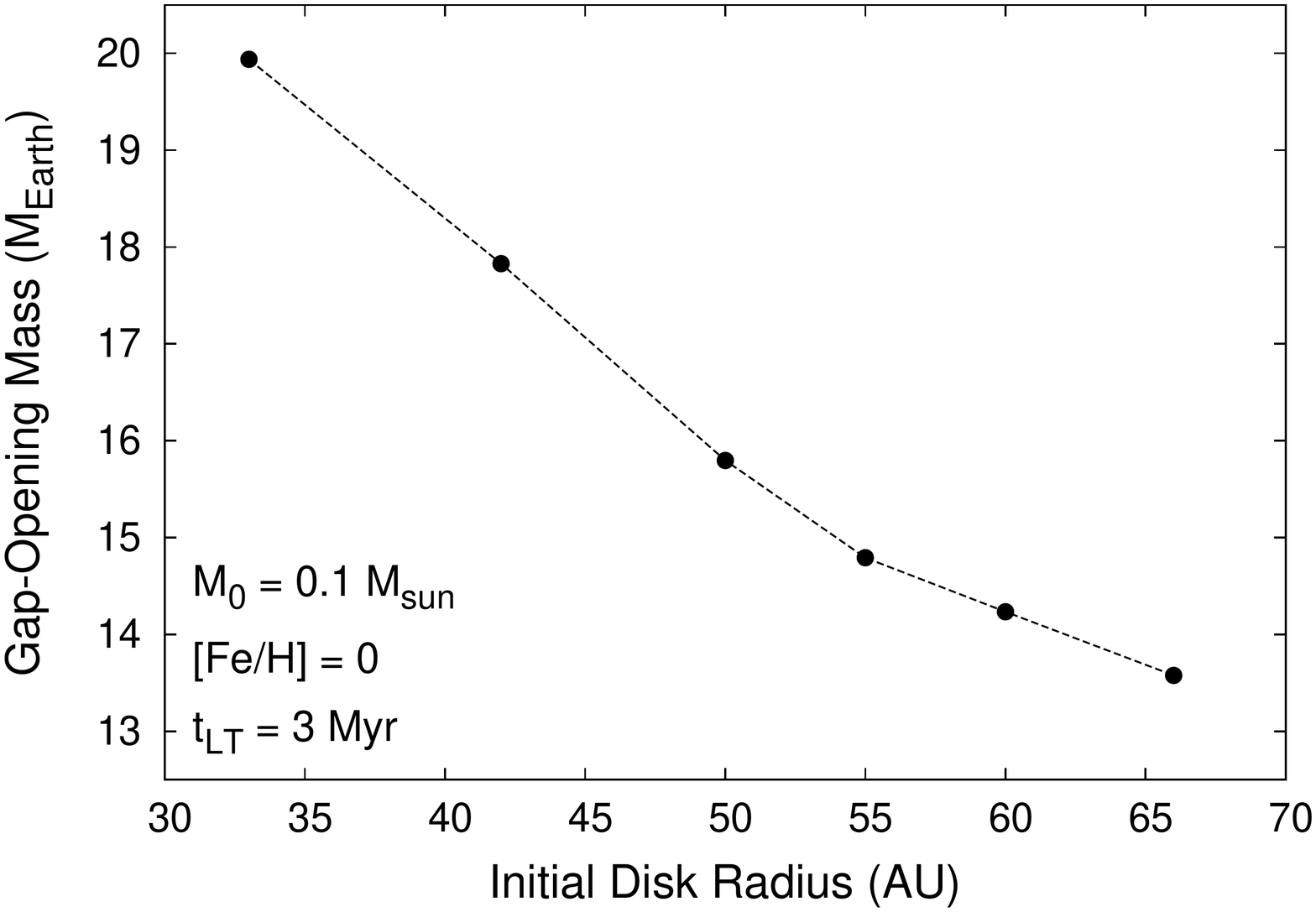}
\caption{We plot gap-opening masses for planets forming at the ice line in disks with different initial radius settings, and otherwise fiducial parameters. The gap-opening mass is systematically larger in more compact disks.}
\label{GapMass}
\end{figure}

In figure \ref{Radius_Compare} (left panel), we consider a series of individual planet formation tracks that consider different initial disk radii, holding other parameters (disk mass, metallicity, and lifetime) constant. This is done to investigate the results we see regarding the formation frequency of gas giants and super Earths at the ice line, and their dependence on the initial disk radius. We highlight that all initial radii settings result in tracks that efficiently form zone 3 gas giants, with the exception of the $R_0$ = 50 AU case that undergoes significant inward migration during its formation. We also note that the initial position of the cores (i.e. the initial position of the ice line) shifts inwards as the disk radius is increased, due to the lower disk surface density.

In figure \ref{Radius_Compare}, right panel, we show the solid accretion timescale for the series of ice line planet formation tracks computed using equation \ref{Solid_Accretion}. The solid accretion rate scaling is $\tau_{\rm{c,acc}} \sim r_p^{3/5} \Sigma_d^{-1} \Sigma_g^{-2/5}$, with the dust and gas surface densities being calculated at the location of the planet (i.e. the ice line). As was discussed in section 2.2, the dust surface density at the ice line increases as the disk radius is increased. The ice line's position does shift inward for larger $R_0$, however this change is small compared to the variations in the disks' extents that we have explored. The dust surface density is larger at the ice line for bigger disk radii settings simply because the drift-limited region is larger and there is more dust from the outer disk that is transported into the ice line. Therefore, both the smaller $r_p$ and larger $\Sigma_d$ contribute to a shorter solid accretion timescale in larger disks. The gas surface density, however, is larger for the smaller disk settings due to the initial disk mass being held constant with the radius changing.

Combining these three effects, the right panel of figure \ref{Radius_Compare} clearly shows that the solid accretion timescale at the ice line is shorter as the disk radius is increased. The difference between the $\tau_{\rm{c,acc}}$ values is largest at early times, and when comparing more compact disks (i.e. the difference is smaller when comparing two large $R_0$ settings). This shows that solid accretion is most efficient at the ice line in large disks. Since subsequent gas accretion is dependent upon the solid accretion stage, this indicates that the ice line should be more efficient at forming gas giants in disks with larger initial radii. We also note that the solid accretion timescales converge within 1 Myr, and by that time the difference in timescales is small across the range of $R_0$ values considered. 

This trend is shown in our population results for disks with initial radii 50 AU and larger (see figure \ref{Zone_Frequencies}). The super Earth formation frequency is maximized at 50 AU, mainly due to formation of this class of planets at the ice line. Beyond 50 AU, the frequency of zone 3 gas giants increases due to faster solid accretion caused by radial drift transporting more solids to the ice line in bigger disks. However, our population results also show the gas giant formation frequency to be large in disks smaller than 50 AU, namely the 33 AU and 42 AU cases. 

This begs an interesting question because as we have seen, more massive planets are expected for large disks based on the amount of solid material is available at the ice line. Therefore a different aspect of our planet formation model must be causing gas giants to form more frequently than super Earths within the ice line in disks with small initial radii.  

Another important aspect of giant planet formation is the amount of gas that will accrete onto them. We note that this is controlled by the gap opening mass. Accordingly, in figure \ref{GapMass}, we plot the gap-opening masses of planets forming at the ice line in disks with different initial radii, computed using equation \ref{GapOpening}. We see immediately that the gap-opening masses for these planets are larger in more compact disks. This trend can be simply explained by considering the gap-opening mass' dependence on the disk aspect ratio, with $M_{\rm{GAP}}\sim h^3$ or $h^{5/2}$, depending on whether the viscous or thermal gap-opening criterion is met.

The disk aspect ratio scales as $h \sim T^{1/2}r_p^{1/2}$ and we note that, since we are considering planet formation to take place at the ice line, the local disk temperature will be the same (the sublimation temperature of water) regardless of the initial disk radius setting. This simplifies the above gap-opening mass scaling to $M_{\rm{GAP}}\sim r_p^{3/2}$ or $r_p^{5/4}$. Since the initial masses of the disks we are comparing are the same, the column density in more compact disks will be higher, and in turn the ice line trap (which sets the planets' radii, $r_p$) will exist at a larger radius.

Thus, larger ice line  planet radii $r_p$ in smaller disks leads to the trend seen in figure \ref{GapMass}, whereby the gap opening masses of planets forming at the ice line are larger in disks with smaller initial radii. As previously mentioned (see initial position of planets in figure \ref{Radius_Compare}, left panel), the position of the ice line does not vary drastically with initial disk radius, but the sensitive scaling of the gap-opening mass with planet radius causes for the somewhat large range of gap-opening masses encountered across the investigated range of $R_0$.

Planets forming at the ice line in more compact disks (the 33 and 42 AU cases) are less impacted by gas accretion termination due to their larger gap-opening masses, and we attribute the high gas giant formation frequency in these smaller $R_0$ disks to this. We recall that termination of gas accretion is set by our $f_{\rm{max}}$ parameter (see equation \ref{MaximumMass}), with low $f_{\rm{max}}$ settings of order unity corresponding to planets whose gas accretion is terminated shortly after they exceed their gap-opening masses. In these cases, low $f_{\rm{max}}$ settings of order unity can terminate gas accretion at an intermediate super Earth - Neptune mass that results in a planet population zone 5. Due to the larger gap-opening mass in smaller disks, these planets have a smaller $f_{\rm{max}}$ range that can lead to planets populating zone 5. This outcome is therefore less likely for smaller settings of $R_0$, but has a larger effect on planets forming in disks with larger $R_0$ settings.

To summarize, the large formation frequency of gas giants at the ice line in small disks is a result of the larger gap-opening masses, and the planets therefore being less subjected to the effects of gas accretion termination. On the large disk radius end of explored $R_0$ range, the high formation frequency of gas giants at the ice line is a result of the higher solid surface densities at the trap, and the correspondingly shorter solid accretion timescales. The $R_0 = 50$ AU setting, as an intermediate disk radius case, is the least optimal setting for formation of gas giants at the ice line (as it does not benefit from either of the effects that help produce gas giants in the small or large disk cases), but is the optimal condition for forming the largest observed planetary population - the Super Earths and Neptunes.

\subsection{Comparison to Constant Dust-to-Gas Ratio Models}

In Appendix C, we do a full comparison between M-a distributions resulting from this paper's models that include dust evolution to models of our previous work (\citet{Alessi2018}) that assume a constant dust-to-gas ratio of $f_{\rm{dtg}}$ = 0.01.

We find that the trade-off between warm Jupiters and super Earths formed at the ice line, depending on the setting of the initial disk radius, is a result only seen when dust evolution is included. Constant dust-to-gas ratio models show significantly less sensitivity to the initial disk radius.

We also find that, when dust evolution is included, our models can produce super Earths with smaller orbital radii (down to $\sim$ 0.03 AU) than when a constant dust-to-gas ratio is assumed. This is caused by radial dust drift and the resulting delayed formation at the X-ray dead zone, whereby solid accretion rates are negligibly small until the trap migrates near or within the ice line.

%% file: Discussion_Conclusion.tex
\section{Discussion}

\subsection{Population synthesis: Host-star and disk parameters}

\subsubsection{The initial disk radius}
The main result from our initial disk radius parameter study is that the largest super Earth population is formed at an initial disk size of 50 AU. This, in addition with the sensitivity of the ratio of warm Jupiters and super Earths formed at the ice line to the initial disk radius show that planet formation is fundamentally linked to protoplanetary disk properties. The population synthesis technique itself assumes this link between the scatter in the planet M-a distribution and disk properties, however we have built upon results of previous works through a separate parameter study of the initial disk radius. By keeping the initial disk radius constant within each population run while varying the disk's mass, lifetime, and metallicity stochastically in our calculations, we have isolated the effect of $R_0$ by changing it between each population run.

As observations indicate low-mass planets to dominate the M-a diagram in terms of frequency, our conclusion is that intermediate disk sizes ($\sim$ 50 AU) produce best-fit populations, as these models produce the largest super Earth population. This is nicely in accord with MHD simulations of disk formation during protostellar collapse which show that, depending on the setting of the mass-to-magnetic flux ratio ($\mu$) of the collapsing region, comparable disk sizes are produced, supporting our results \citep{Masson2016}. Additionally, this result is supported by the recent observations that show small to intermediate disk sizes to be common (i.e. \citet{Barenfeld2017, Cox2017, Long2019}). 

As the distribution of protoplanetary disk radii becomes better constrained by observations, this can be incorporated into our population synthesis models as an additional parameter that is stochastically varied in each population run. In this work, we did not include any correlation between disk masses and radii to isolate the effect of changing the disk radius on outcomes of our planet formation model. Such a correlation has been shown to exist, indicated by the correlation between dust continuum fluxes and either dust disk radii \citep{Tazzari2017, Tripathi2017} or gas disk radii \citep{Ansdell2018}. Again, as observations better constrain these disk properties, changes in these disk parameters' distributions, and any correlations among them, can be readily incorporated in our population synthesis models. Updating these distributions as more data becomes available will be important, since the resulting M-a planet distributions are to a large degree shaped by disk properties.

We note that, while the investigated range of initial disk radii clearly has a large effect on the outcomes of planet formation, it is unlikely that the observed range of dust disk radii can be reproduced with the dust model considered in this work. After $\gtrsim$ 1 Myr of evolution (a typical age of an observed disk), the dust distribution exists entirely within the ice line due to the dust model's efficient radial drift. Thus, the range of solid disk radii (spanning the relatively small range in ice line radii) will not reflect the range of initial disk radii investigated. We expect that a means of maintaining a more extended dust distribution, either through reducing the efficiency of radial drift or with the inclusion of dust traps, would lead to a larger range of dust disk radii in evolved disks (see also section 4.6).

However, we emphasize that the earliest stages of disk evolution, where differences in disk conditions for different initial radii are most pronounced, are most crucial for planet formation. This is particularly true for the ice line, where planet formation is seen to depend on $R_0$ most sensitively (see timestamps in figure \ref{Tracks}, left panel).

\subsubsection{Host-star mass}
In this work, we modelled our disks to exist around pre-main sequence G-type stars, and did not explore other spectral classes. In doing so, we were focusing on effects that the disk itself has on outcomes of planet formation as opposed to the host-stellar mass and luminosity. Additionally, by restricting our models to Solar-type stars we are comparing with the majority of the exoplanetary data. 

Previous works have shown that the stellar mass also plays an important role in the outcomes of planet formation. \citet{IdaLin2005} showed that the stellar mass affects the ratio of short-period gas giants to Neptune-mass planets. Additionally, the results of \citet{Alibert2011} show that the scaling of disk properties (lifetime and mass) with stellar mass is an important inclusion and greatly influences the final outcomes of population synthesis calculations. Including variation in host-stellar mass is a prospect for future work, and building off of the results of \citet{Alibert2011}, including host-stellar mass-dependent distributions of disk lifetimes, masses, metallicities, \emph{and disk radii} will be important to fully explore the effects of stellar mass on outcomes of our population synthesis models. It is currently unlikely that sufficient observational data exists to correlate all of these disk properties' distributions with host-stellar mass.

\subsection{Implications for super Earth compositions}

This work's optimized model of $R_0$ = 50 AU resulted in the largest population of zone 5 planets. In this model, the super Earth population consists almost entirely of planets formed at the ice line and at the dead zone. This has implications for these planets' compositions. Planets formed at the ice line will have a significant fraction of their solid mass in ice, while planets formed in the dead zone trap will have nearly no ice accreted, since all of their solid accretion takes place within the ice line. Super Earth compositions are therefore bimodal in these models. Additionally, we find that the super Earths with larger orbital radii ($\gtrsim$ 1 AU) are predominantly ice line planets, and those at smaller orbits were formed at the dead zone. Our best-fit model therefore predicts a jump in the mean density of super Earth solid cores at $\sim$ 1 AU, transitioning from dry, dense planets formed at the dead zone to those with a substantial ice fraction formed at the ice line. We will follow up on this issue in considerable detail in our next paper.

At larger initial disk radii ($R_0$ = 66 AU), we find that zone 5 is nearly entirely populated by planets formed at the dead zone and heat transition, with the ice line mainly forming warm Jupiters. In this case, the bimodality of the super Earth compositions will be lost, since in both cases of planets forming in the dead zone and heat transition traps, solid accretion will be restricted to take place within the ice line. This is due to radial drift efficiently depleting the outer disk of solids, and therefore it is not until the traps migrate within the ice line that planets forming at either the heat transition or dead zone are able to accrete significant amounts of solids. In this case there would be no transition among the core compositions (or densities) in super Earths, despite there being a clear transition between short period super Earths formed mostly in the dead zone, with super Earth on longer orbits being formed in the heat transition. 

\subsection{Low solid accretion in outer disk \& additional planet traps}

An additional implication of low solid accretion rates in the outer disk due to radial drift is that outer planet traps not included in our model would contribute planets in the observable region of the M-a diagram. Traps such as additional condensation fronts (such as CO$_2$ \citep{CPA2019}), or resonances of traps we include in our model would exist outside of the ice line for the entirety of disk evolution. Since planets forming in these traps would be accreting from the drift limited region of the disk, there would be minimal solid accretion, and minimal growth of planetary cores. Planet formation at these traps would therefore only result in very low mass failed cores (comparable to planet formation in the heat transition in the $R_0$ = 33 AU case), and would not contribute even to the zone 5 population. Our results remain unaffected regardless of whether or not additional traps in the outer disk are included, justifying their omission.

\subsection{Increasing the short-period super Earth population}

While the best-fit model produced the largest super Earth population, the formation frequencies of zone 5 planets in our models is still not large enough to compare with the data. The inclusion of the dust model results in more short-period super Earths being produced (down to orbits of $\sim$ 0.03 AU), with the majority of super Earths formed in our model having orbits between $\sim$ 0.1-3 AU. Similar to our previous work \citep{Alessi2018}, we again find that our models produce many super Earths between 1-3 AU, and thus predict many low-mass planets to exist just outside the $\sim$ 1 AU outer limit where super Earths have been detected via transits.

The observed M-a diagram shows the existence of more super Earths between 0.01-0.1 AU than our model produces. However, with observational biases accounted for, the occurrence rate study of \citet{Petigura2018} shows that super Earth and Neptune-mass planets' frequencies peak just within 0.1 AU, with occurrence rates decreasing at smaller orbital radii. Our results compare well with this data as the low orbital radius end of the bulk of our super Earth populations lie at $\sim$ 0.1 AU. 

At yet smaller orbital radii, our populations do not compare well with the observed M-a diagram or occurrence rate studies due to the lack of super Earths at orbits $<$ 0.1 AU in the cases of our best-fit model with $R_0$ = 50 AU or larger. An exception is the smallest initial disk radius case of $R_0$ = 33 AU where the low orbital radius end of the super Earth population extends down to 0.05 AU.


With the inclusion of dust evolution, our models show the core accretion model is capable of producing short-period super Earths reliably down to 0.05 AU. Nonetheless, we identify three mechanisms by which the very short-period super Earth population (0.01-0.1 AU) could be increased in our calculations to better compare with the data. 

\subsubsection{Planet-planet dynamics}

Firstly, we assume our planetary cores form in isolation and neglect any dynamics effects. Post-disk dynamics can have an effect on the final orbits of planets formed during the disk phase in our calculations, as was shown in \citet{Ida2013}. Planet-planet scattering can reduce the orbital radius of the remaining planet by up to a factor of two - the case for scattering between two equally massive planets. We therefore do not expect this to have a drastic effect on our planet populations, although we do note this as a means by which planets' orbital radii can be reduced. Investigating the ways in which dynamics can affect our calculations during and after the disk phase remains a prospect for future work. We highlight that our models form many low-mass ($<$ 1 M$_\oplus$) planets that can take place in accretion or scattering if dynamics was included during the post-disk phase.

\subsubsection{Corotation torque saturation}

As discussed in Appendix B, saturation of the corotation torque prior to gap opening and type-II migration is another method by which more short-period planets could be formed. Here, we only include the trapped type-I migration phase following the results of \citet{Alessi2017}, using the timescale approach of \citet{Dittkrist2014} to determine if a saturated type-I migration phase applies. We note that the gap-opening mass and the mass at which the corotation torque saturates are comparable and sensitive to model parameters. 

As was noted in \citet{Hasegawa2016}, if the corotation torque saturates prior to gap-opening, a saturated type-I migration regime would apply as an intermediate step between trapped type-I migration and type-II migration (the two regimes included in this work). This would remove planets from their traps prior to them reaching their gap-opening masses, and planet-induced gaps observed in disk dust distributions may not have to align with planet traps (or condensation fronts). If a saturated corotation torque phase applied prior to the onset of type-II migration in our model, then the orbital radii of planets would indeed be smaller, and could lead to a reduction in the orbital radii of formed super Earths. This would also, however, lead to more planets being accreted onto the host star.

\subsubsection{The embryo assembly mechanism}

Lastly, as suggested in \citet{Hasegawa2016}, the embryo assembly method of forming super Earths could lead to more short-period super Earths beyond what our models are capable of producing. This is an alternate scenario to the core accretion mechanism, whereby planetary embryos migrate to the inner edge of the disk but do not accrete gas (due to their low masses), and undergo collisions after the disk phase to build up a super Earth. With the inclusion of dust evolution, however, the core accretion model is better able to produce short-period super Earths, so we speculate that a change in a model detail within the core accretion approach could lead to more super Earths in the 0.01-0.1 AU range as opposed to requiring a different formation mechanism entirely.

\subsection{Zone 1 \& Zone 2 Populations}

Our populations produce too large a fraction of gas giants, particularly in zones 1 \& 2, when compared with data from occurrence rate studies \citep{Santerne2016, Petigura2018}. Many of the planets in zones 1 \& 2 are formed within the dead zone trap. We recall that dead zone planets only begin accreting appreciable amounts of solids once they have migrated within the ice line, a consequence of radial drift removing solids from the outer disk. Thus, the over-production of short-period gas giants from the dead zone is another result that can be attributed to the efficiency of radial drift in the dust model. The resulting high surface densities of solids in the inner disk lead to efficient solid accretion onto dead zone planets once they have migrated within the ice line, leading to many short-period gas giants. Additionally, an increased super Earth population (discussed in the previous subsection 4.4) would result in a comparatively smaller frequency across all gas giant zones, so the over-production of gas giants is related to the under-production of super Earths.

Our models do not show a separation between hot Jupiter and warm Jupiter populations, regardless of the setting of the initial disk radius. We therefore do not reproduce the reduced frequency of period-valley giants at orbital periods $\sim$ 10 days seen in occurrence rate studies \citep{Santerne2016, Petigura2018}. We note that this range of reduced occurrence rates for the period-valley giants is much smaller than the raw exoplanet data on the M-a diagram would indicate (across the extent of zone 2 as indicated by \citet{ChiangLaughlin2013}).

The best fit model from our previous paper in this series \citep{Alessi2018} resulted in a large population of warm Jupiters as well as a clear separation between warm Jupiters and shorter period hot Jupiters (see figure \ref{Pop_Compare}, top right panel), reproducing this feature of the data. However, we recall that our previous work did not account for any dust evolution effects. Additionally, this separation is only seen in constant dust-to-gas ratio models when an initial disk radius of $R_0$ = 33 AU is used.

\subsection{Efficiency of radial drift \& dust trapping}

The planet formation results of this paper are influenced to a large degree by the dust evolution model, and particularly the efficient radial drift that transports solids outside of the ice line inwards. We note that radial drift in the \citet{Birnstiel2010} (for which the dust model used in this work \citep{Birnstiel2012} is a numerical fit) was found to be too efficient when compared with spectral energy indices of observed disks \citep{Birnstiel2010b}. Additionally, the offset in gas and dust dust radii can by explained by differences in optical depths for the majority of cases, and only require invoking radial drift for the most extreme discrepancies \citep{Facchini2017, Facchini2019}. In our calculations, high dust-to-gas ratios are maintained only within the ice line ($\sim$ 5 AU) even at early stages in the disk's evolution due to efficient radial drift in the outer disk. Comparing the ice line radius to the extent of the disk ($\sim$ 50 AU), there is indeed an extreme discrepancy between the dust and gas disk radii in our models as a result of radial drift in the \citet{Birnstiel2012} model being too efficient. We found the dust model to be largely insensitive to the fragmentation and drift parameters ($f_f$ and $f_d$; see section 2.2) used to fit the \citet{Birnstiel2012} simplified two-population model to the full numerical calculation of \citet{Birnstiel2010}. The dust-to-gas ratio profiles and the rates of radial drift remained mostly unaffected through a large variation in each parameter.

As discussed in \citet{Pinilla2012}, dust trapping at local pressure maxima is a means of maintaining extended dust distributions despite efficient radial drift elsewhere in the disk. This mechanism is consistent with disk observations that show local structures in dust (i.e. \citet{Casassus2015}) indicating dust trapping. Including dust traps in our model as a physical means of slowing radial drift would likely affect our results, as the solid accretion phase is sensitive to the distribution of solids throughout the disk. If dust traps were able to maintain high solid surface densities in the outer disk, and prevent solids from quickly radially drifting towards the ice line, planet formation timescales at all the traps (particularly the heat transition and the ice line) would be impacted. Additionally, if the dust traps were also a location of a planet trap, the local enhancement of solids would lead to efficient planet formation at the dust trap. This is similar to the behaviour at the ice line in our models. Although it is not modelled as a dust trap, there is a local enhancement of solids at the ice line during early stages of disk evolution due to the changing fragmentation velocity across the ice line's radial extent, leading to efficient solid accretion and planet formation at the ice line. 

Prior to planet formation taking place, it is unlikely that dust traps could exist at arbitrary locations in the disk as opposed to existing at inhomogeneities and local disk structures - namely planet traps. At early times in the disk's evolution, both the heat transition and outer dead zone radii exist outside the ice line, as do condensation fronts of volatiles other than water, such as CO$_2$. If these, in addition to the water ice line, were all treated as dust traps in the model, the dust surface density would be larger over a more extended range of radii, and there would be a smaller discrepancy between the gas and dust radii despite radial drift being present in the calculations. Dust traps were not included in this work as one of our main goals was to explore the unhindered effects of radial drift on our planet formation models. 

It is interesting to consider the populations computed in this work that include radial drift, and those resulting from the constant dust-to-gas ratio assumption of \citet{Alessi2018} as two extremes in treatment of radial drift. In this work's case, radial drift is too efficient, while it is ``turned off'' when neglecting radial drift effects. Therefore, the populations of this work and our last can be thought of as bracketing the true effects of radial drift on planet formation (in which dust trapping would need to be accounted for). We thus identify dust trapping as an important inclusion in models that include dust evolution and radial drift, and incorporating this into our dust treatment is a prospect for future work.

\subsection{MHD disk winds vs. turbulent alpha}

While the turbulent $\alpha_{\rm{turb}}$ setting used in this work is within the accepted range based on observed line widths in disks \citep{Flaherty2018}, it remains possible that the low levels of turbulence observed in disks can be attributed to them evolving through MHD-driven disk winds as opposed to MRI-turbulence, as the \citet{Chambers2009} model used in our calculations assumes. A key difference between the two mechanisms of angular momentum exchange is the disks evolving via MRI-turbulence spread to conserve angular momentum, while winds-driven models do not as their angular momentum is carried in the wind-driven material \citep{Pudritz2019}. 

Changing the disk model to one that evolves through MHD winds could certainly affect our results, but most crucially within the region of the disk where planet formation takes place, which is confined to occur within the ice line with the current treatment of radial drift, but is generally $\lesssim$ 10 AU in planetesimal accretion models. While the surface density evolution in the outer disk would be different between the two mechanisms of disk evolution due to spreading in the case of MRI-turbulence, this alone would not greatly influence planet formation results. 

If an MHD winds-driven model were used, the different surface density profile in the inner disk would, however, affect various stages of our planet formation model. The lower level of turbulence throughout the disk despite similar overall $\alpha$ settings would affect dust growth and radial drift rates. The set of planet traps we include would also change, as winds-evolving disk models show local maxima in surface density profiles (i.e. \citet{Ogihara2018}). Conversely, the outer edge of the dead zone may be removed as a trap due to the overall lower levels of $\alpha_{\rm{turb}}$. It has additionally been shown that the co-rotation torque works very differently in inviscid MHD wind-driven disks \citep{McNally2017, McNally2018, Kimmig2019}. Ultimately, all of these aspects combined might affect our population synthesis results, and incorporating a winds-evolving disk model will be the focus of our future work in this series.

\subsection{Gas accretion termination}

Our treatment of the late stages of planet formation only considers gas accretion to proceed at the Kelvin-Helmholtz rate prior to being terminated artificially when planet's reach their maximum mass, set by the $f_{\rm{max}}$ parameter that is varied in our population synthesis models. While terminating accretion in such a manner is a simplified approach, stochastically varying the $f_{\rm{max}}$ parameter in our populations results in a range of gas giant masses that is comparable with the data.

 \citet{Lambrechts2019} find that even massive planets can maintain high accretion rates and there is no self-driven mechanism to halt planetary accretion. They found their results were unaffected by gap-formation due to the large amount of material flowing through the gap available for accretion onto the planet \citep{Morbidelli2014}. This result supports our treatment of late stages of gas accretion, in the sense that gas accretion is unhindered prior to the planet reach its maximum mass.

There are two alternate treatments of truncating accretion for high mass planets. The first is the disk-limited accretion mechanism (i.e. \citet{Tanigawa2016}) whereby the reduced accretion rate through the disk itself truncates accretion onto the planet. Second is the magnetic termination of gas accretion \citep{Batygin2018, Cridland2018} whereby the interaction of the planet's magnetic field and the disk results in an accretion cross-section that inversely scales with planet mass, leading to termination of accretion at high masses. 

In both of these alternate treatments, we argue that the planet's final mass is still ultimately set by a model parameter\footnote{In disk-limited accretion, the fraction of the disk accretion rate that accretes onto the planet is parameterized, setting the planet's final mass. In the case of magnetic termination, the forming planet's magnetic field strength is a parameter that sets the final mass of the planet.}, and in that regard do not improve over our $f_{\rm{max}}$ approach. If we instead were to use an alternate approach of terminating accretion, we do not expect our final populations to be affected, as a suitable range of model parameters would need to be chosen (as is the case with $f_{\rm{max}}$) to obtain a reasonable range of gas giant masses. 

We note that during the runaway growth phase the Kelvin-Helmholtz accretion rate is systematically higher than the disk-limited accretion rate that has been used in many previous works (i.e. \citet{Machida2010, Dittkrist2014, Bitsch2015}). Additionally, \citet{Hasegawa2019} found that including a disk-limited accretion phase is necessary to reproduce the exoplanetary heavy-element content trend. If we were to include the disk-limited accretion phase, we expect that our giant planet populations would systematically lie at smaller orbital radii. However, we do not expect this shift to be extreme, and depending on model parameters (such as the planet's envelope opacity, and the fraction of material accreted through the disk that accretes onto the planet), both methods can lead to quite similar results.  

\section{Conclusions}

In this work, we have examined the role of the initial disk radius in core accretion models through a comprehensive set of planet population synthesis calculations. We have also updated our calculations in their treatment of dust to a physical model that combines dust growth, fragmentation, and radial drift. Including dust evolution effects has shown a drastic change in the population results shown in \citet{Alessi2018} that assumed a constant dust-to-gas ratio of $f_{\rm{dtg}}$ = 0.01.

Our major finding - that intermediate disk radii of the order of 50 AU drives the appearance of the observed M-a diagram - indicates that planet formation and star formation processes are intimately linked.   The characteristic disk radius is determined by a combination of gravitational collapse of turbulent regions and their angular momentum evolution due to magnetic braking and outflows.   

We list our main conclusions below:
\begin{itemize}
\item \emph{The ice line is the most important location for warm Jupiter formation.} The trap becomes locally enhanced in solids early in the disk's evolution due to radial drift removing solids from the outer disk and transporting them towards the ice line. This effect restricts the region of the disk where solid accretion can take place at an appreciable rate to within the ice line, as solid accretion in the outer disk is inefficient due to the low solid surface densities. This has the largest affect on planet formation in the heat transition trap in our models.
\item \emph{Planet formation is fundamentally linked to the characteristic initial radius of the protoplanetary disk population}. The ratio of super Earths to warm Jupiters formed at the ice line is sensitive to the setting of the initial disk radius. The smallest (33 AU) and largest (66 AU) initial disk sizes (which would form in collapsing regions with strong magnetic fields, and pure hydrodynamics collapse, respectively) resulted in the ice line producing many more warm Jupiters than super Earths
\item \emph{An initial gas disk size of 50 AU produces the largest super Earth population.} This is a feature of planet formation at the ice line, for which we find that super Earth formation is optimized for intermediate disk radii settings (that would form from collapsing regions with moderate magnetic field strengths). Gas giant formation at the ice line is more efficient (1) at smaller disk radii due to larger gap-opening masses and planets being less effected by related gas accretion termination; and (2) at larger disk radii due to larger solid surface densities at the ice line trap and correspondingly shorter solid accretion timescales. These effects minimize at the intermediate disk radius of 50 AU.
\item \emph{Inclusion of radial drift is essential to form short period super Earths.} Notably, planet formation at the dead zone trap, which is delayed until the dead zone migrates to within the ice line, results in super Earth formation with orbital radii as small as 0.03 AU. Our previous treatment \citep{Alessi2018} that neglected dust evolution effects and assumed a constant dust-to-gas ratio throughout the disk was unable to form super Earths with orbital radii significantly less than 1 AU.
\end{itemize}

In our upcoming work in this series, we will investigate the chemical compositions of these various populations using planets produced in our populations and the disk chemistry model of \citet{Alessi2017}. We will combine our computed planet compositions with an interior structure model to examine our populations' distributions on the mass-radius diagram. In future work, we will consider the effects of a MHD-winds driven disk model on our populations, as an alternative to the MRI-turbulence driven model we have thus far considered. 

%% file: AppendixA.tex
\section{Protoplanetary Disk Model}

We use the \citet{Chambers2009} 1+1D semi-analytic model to calculate disk structure and evolution. This model calculates self-similar solutions to the disk evolution equation,
\begin{equation} \frac{\partial \Sigma}{\partial t} = \frac{3}{r}\frac{\partial}{\partial r}\left[r^{1/2}\frac{\partial}{\partial r}\left(r^{1/2} \nu \Sigma\right) \right]\, , \label{ViscousDisk} \end{equation}
where $\Sigma(r,t)$ is the disk's evolving surface density profile, and $\nu(r,t)$ is the disk's viscosity. Self-similar solutions to equation \ref{ViscousDisk} can be obtained by parameterizing the disk's viscosity using an effective viscosity coefficient $\alpha$ \citep{SS1973, LBP1974},
\begin{equation} \nu = \alpha c_s H \, , \label{SS_Viscosity} \end{equation}
where $c_s$ is the disk sound speed and $H$ is the disk scale height.

\begin{figure*}
\centering
\includegraphics[width = 2.2in]{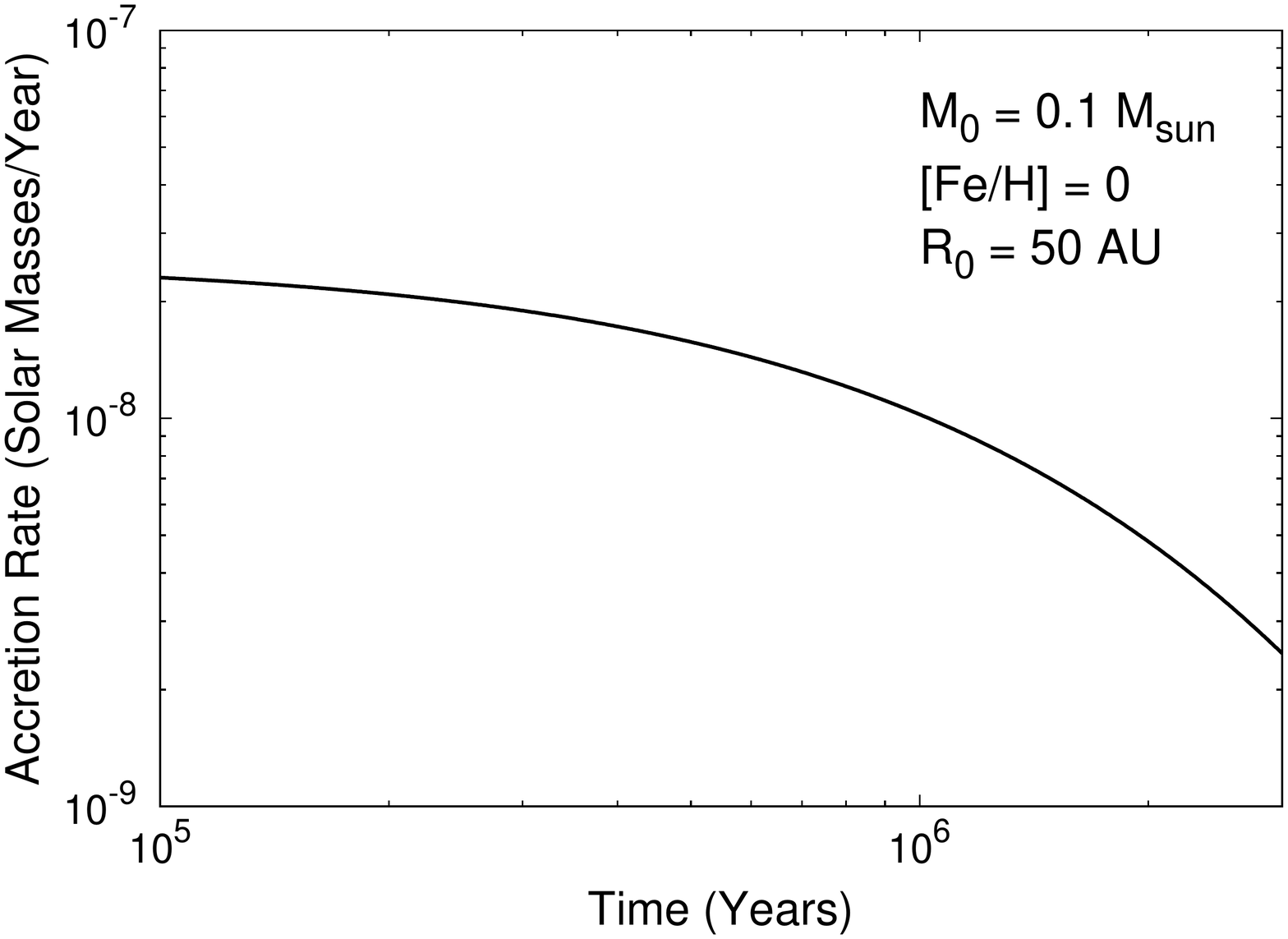} \includegraphics[width = 2.2 in]{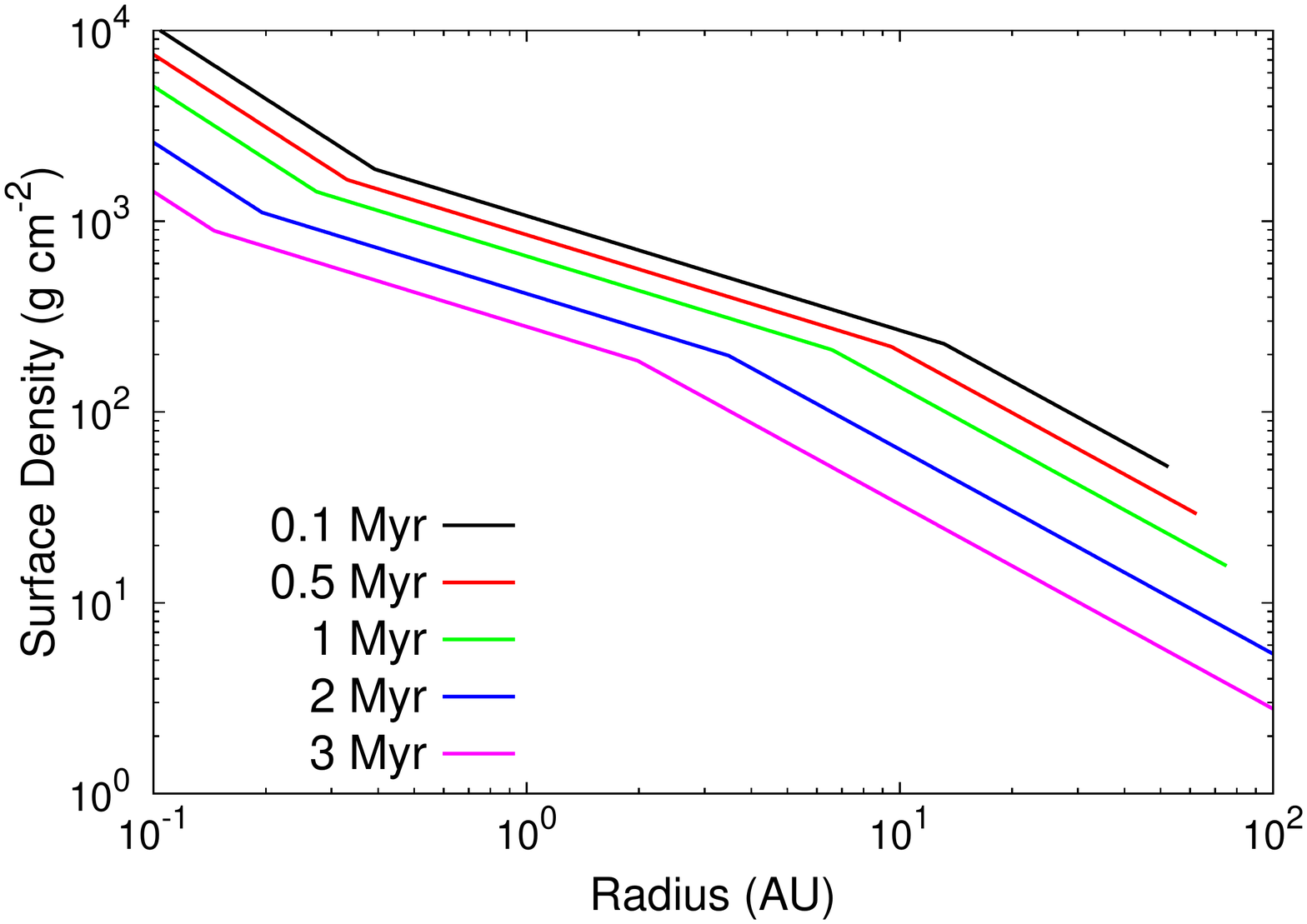} \includegraphics[width = 2.2 in]{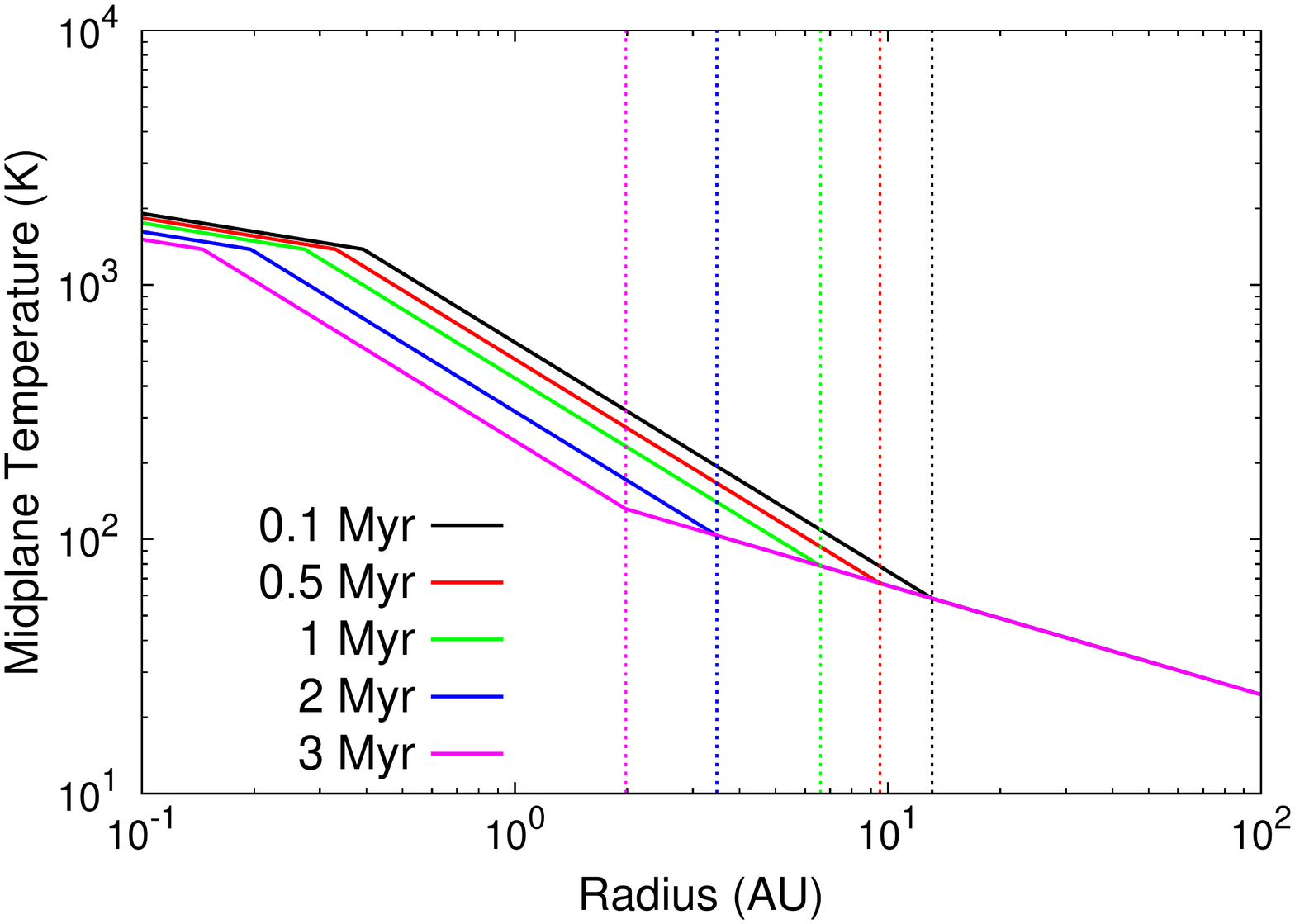}
\caption{The time evolution of the accretion rate (left) and radial profiles of surface density (middle) and midplane temperature (right) are plotted for a fiducial disk. The vertical dashed lines in the right panel mark the location of the heat transition at the boundary of the viscously- and radiatively-heated regimes.}
\label{DiskPlot}
\end{figure*}

A globally-constant value of $\alpha$ is required in order to obtain self-similar solutions to equation \ref{ViscousDisk}. Angular momentum transport in protoplanetary disks can take place through either MRI-turbulence or MHD-driven disk winds. Thus, $\alpha$ can be written as a sum of the effective viscosities of each source of angular momentum transport ($\alpha_{\rm{turb}}$ and $\alpha_{\rm{wind}}$, respectively),
\begin{equation} \alpha = \alpha_{\rm{turb}} + \alpha_{\rm{wind}} \, . \label{EffectiveAlpha} \end{equation}
In this work, we set $\alpha = 10^{-3}$ in all calculations. This setting is consistent with the upper limit of $\alpha_{\rm{turb}} < 0.007$ measured in the TW-Hya disk \citep{Flaherty2018}. 

MRI-driven turbulence requires the disk to have a critical ionization fraction in order to operate. The high-density inner region of a disk prevents ionizing radiation from reaching the midplane, preventing MRI-turbulence (the disk's \emph{dead zone}). Previous works considering only the MRI-turbulence contribution to angular momentum transport, such as \citet{Gammie1996} and \citet{MP2003} have shown that within the disk's dead zone, $\alpha_{\rm{turb}} \sim 10^{-5} - 10^{-4}$, while in the outer, turbulently active region of the disk, $\alpha_{\rm{turb}} \sim 10^{-3} - 10^{-2}$. However, MHD-driven disk winds have been shown to maintain accretion rates within the disk's dead zone \citep{BaiStone2013, Gressel2015, Gressel2015b, Bai2016}. Our assumption of a globally constant effective $\alpha$, despite a radially changing $\alpha_{\rm{turb}}$, is consistent with these results - requiring an $\alpha_{\rm{wind}}$ contribution to maintain a radially constant accretion rate.

In all calculations, we consider a pre-main sequence G-type host star, with mass 1 M$_\odot$, radius 3 R$_\odot$, and an effective temperature of 4200 K. Therefore, any variations in planet populations that would result from different stellar properties (i.e. \citet*{IdaLin2005, Alibert2011}) are not included in this work. Instead, we focus on the effects that disk properties have on the planetary M-a distribution. We use a fiducial initial disk radius of $R_0 = 50$ AU, and recall that disks will viscously spread as they evolve via MRI-turbulence. 

Disk evolution takes place in our model through the combined effects of viscous accretion and photoevaporation. The latter effect is caused by high-energy radiation from the host star (UV and X-rays) that continuously disperse disk material \citep{Pascucci2009}. We model the time-evolution of the disk accretion rate to be,
\begin{equation} \dot{M}(t) = \frac{\dot{M}_0}{(1 + t/\tau_{\rm{vis}})^{19/16}} \exp \left(-\frac{t - t_{\rm{int}}}{t_{\rm{LT}}}\right) \;, \label{AccretionRate} \end{equation}
where $\tau_{\rm{vis}}$ is the viscous timescale, $\dot{M}_0$ is the initial accretion rate at $t_{\rm{int}} = 10^5$ years, and $t_{\rm{LT}}$ is the disk's lifetime. Equation \ref{AccretionRate} includes an exponential photoevaporation factor multiplying the viscous accretion rate evolution of \citet{Chambers2009}. At early stages of the disk's evolution, photoevaporation is a small modification on viscous accretion, but rapidly disperses the disk (on 10$^4$ year timescales, short compared to the $\sim 10^6$ year viscous timescale) once the photoevaporative rate becomes comparable to the viscous accretion rate \citep*{Owen2011, Haworth2016}. We therefore assume the disk to rapidly clear at $t=t_{\rm{LT}}$, ceasing planet formation and migration. 

We assume a constant disk opacity, with metallicity scaling, of \citep{Chambers2009, Remy2014},
\begin{equation} \kappa = 10^{[\rm{Fe}/\rm{H}]} (3 \, \rm{cm}^2 \,\rm{g}^{-1}) \, . \end{equation}
That is, the disk opacity has no radial or temporal variations. The exception to this is in the innermost `evaporative' region, $r_e \lesssim 0.3$ AU, of the disk where the temperature exceeds 1380 K and dust grains sublimate. Here the opacity is modified to \citep{Stepinski1998},
\begin{equation} \kappa = 10^{[\rm{Fe}/\rm{H}]} \left(\frac{T}{1380\,\rm{K}}\right)^{-14} (3 \, \rm{cm}^2 \,\rm{g}^{-1}) \, , \end{equation}
where $T$ is the midplane temperature.

By assuming a radially-constant opacity over the majority of the disk's extent, we are neglecting the opacity variation that would arise at condensation fronts - the physical cause for planet trapping at the ice line. This is not a necessary inclusion in our model, however, as we do not directly compute planet-disk torques when modelling planet migration in this work. 

Our disk model can be divided into three regions: an outer region heated by radiation from the host star, an inner region heated by the generalized viscous heating, and the innermost `evaporative' region, within the viscously-heated regime where the dust opacity is modified due to grains sublimating. The heat transition, $r_t$, separates the two heating regimes and is a planet trap in our model. We note that within the disk's dead zone (where angular momentum is transported via disk winds), the generalized viscous heating is due to Ohmic dissipation at the midplane (a non-ideal MHD heating effect). 
\begin{table}
\caption{The accretion rate and radius scalings of surface density ($\Sigma$) and midplane temperature ($T$) in the three regions in the disk model.}
\centering
\begin{tabular} {|c|c|c|}
\hline
$r<r_e$ & $r_e < r < r_t$ & $r > r_t$ \\
\hline
$\Sigma \sim \dot{M}^{17/19}r^{-24/19}$&$\Sigma \sim \dot{M}^{3/5} r^{-3/5}$ & $\Sigma \sim \dot{M} r^{-15/14}$ \\
$T \sim \dot{M}^{2/19}r^{-9/38}$& $T \sim  \dot{M}^{2/5} r^{-9/10}$ & $T \sim r^{-3/7}$\\
\hline
\end{tabular}
\label{ChambersTable}
\end{table}

In figure \ref{DiskPlot}, we plot the evolution of the disk accretion rate, as well as radial surface density and midplane temperature profiles at various times throughout the evolution of our fiducial disk model. The heat transition trap is seen to shift inwards as the disk evolves. In table \ref{ChambersTable}, we show the radial and accretion rate scalings of surface density and midplane temperature in the three regions of the disk model. 

%% file: AppendixB.tex
\section{Planet Migration \& Formation}
In the type-I migration regime, applying to low-mass planets ($\lesssim 10$ M$_\oplus$), the summed contributions of torques arising at the Lindblad resonances and the planet's corotation region need to be accounted for to compute the resulting planet migration rate. The Lindblad and corotation torques are dependent on the planet's mass and the local disk conditions - namely the power law index of the local surface density and temperature profiles \citep{Paardekooper2010}. For typical disk surface density and temperature profiles, the Lindblad torque on forming planets is negative, and can lead to planetary cores migrating into their host stars on short $\sim 10^5$ year timescales if not counteracted. The corotation torque, a positive torque for typical disk structures, can slow or reverse the migration rate resulting from the Lindblad torque. 

However, in addition to the magnitude of the corotation torque, its \emph{operation} is also sensitive to the disk's local structure \citep{Masset2001, Masset2002}. The libration timescale of material within the corotation region undergoing horseshoe orbits must be shorter than the disk's local viscous timescale in order for the corotation torque to operate \citep{HellaryNelson2012, Dittkrist2014}. If the reverse is true, the librating disk material in the corotation region will not produce a net torque on the planet. In this case, the Lindblad torque (and resulting short inward migration timescale) will operate unopposed.

The corotation torque remains unsaturated and counteracts the Lindblad torque near inhomogeneities and transitions in disks \citep{Masset2006, Sandor2011}. These regions, where the positive corotation torque balances the negative Lindblad torque, are radii of net torque equilibrium, referred to as \emph{planet traps}. Numerical works, such as \citet{Lyra2010} and \citet{Coleman2016} have calculated the sense of migration of orbits near planet traps and have shown traps to be stable equilibria, and as such nearby orbits will migrate towards planet traps. Inhomogeneities and transitions in disks are therefore likely sites of planet formation. Trapped planets will form within planet traps, which themselves migrate inwards on timescales comparable to the disk's evolution time.

Other works that have computed type-I torques on a range of planet masses and disk radii for various disk models (so-called migration maps) have shown zero net torque locations to be common \citep{HellaryNelson2012, Baillie2016, Coleman2016, CPA2019}. However, these works find that the locations of the equilibrium points have mass dependences. In particular, it has been shown that torque equilibrium points do not exist for low mass planets ($\lesssim$ 1 M$_\oplus$). We do not account for mass-dependence of planet traps in our model, and assume planets to be trapped for the entirety of the type-I migration phase - from an initial mass of 0.01 M$_\oplus$ up until the planet opens a gap and transitions to the type-II migration regime. While we note the discrepancy between our treatment of the migration of low-mass planetary cores with these other works, \citet{Coleman2016b} showed traps related to disk inhomogeneities to be mass-independent, which is consistent with our model's treatment.

\citet{Dittkrist2014} considered multiple type-I migration regimes, and found that in many cases, the corotation torque would saturate prior to planets entering the type-II migration phase. This is also in contrast with our assumed mass-independent traps, since we assume the traps (and therefore corotation torque) operate until the planets reach their gap opening masses. If corotation torques were to saturate prior to gap-opening, this would have a large effect on results of our planet formation runs, since the migration rates would be large for planets on the more massive end of the type-I migration regime acted upon solely by the Lindblad torque. However, \citet{Hasegawa2016} showed that the mass at which the corotation torque saturates is comparable to the gap-opening mass. This result was also found in \citet{Alessi2017}, where we additionally show that planets in our model enter the type-II migration regime prior to the corotation torque saturating.

Since planets remain trapped for the entirety of the type-I migration phase, our treatment of type-I migration involves determining the location of the planet traps themselves as these are the locations of planet formation in our model. The traps we include in our model are the ice line, heat transition, and outer edge of the dead zone. This is not an exhaustive list, as planet traps may exist in the inner regions of the disk such as at the inner edge of the dead zone \citep{Gammie1996}, near the silicate sublimation front \citep{Flock2019}, or at the inner edge of the disk itself. It is unclear if the high temperatures ($\gtrsim 1000$ K) of the inner disk would favour planet formation, as solid surface densities would be low due to dust evaporation, and high gas temperatures would hinder gas accretion onto planet cores. These traps may, however, be important to prevent cores that have already formed from being accreted onto the host stars.

Volatile transitions in the cooler regions of the outer disk (i.e. CO$_2$) are an additional set of traps that are not included in our model. We note that \citet{CPA2019} showed that the CO ice line does not trap planets due to the shallow temperature (and thus, opacity) gradient in the outer disk. This work did show that the CO$_2$ ice line can trap planets at larger radii ($\gtrsim$ 20 AU). However, since it exists in the radiatively heated regime of the disk, the trap will not migrate inwards, and will remain at large radii where solid surface densities and accretion rates are small. It is therefore unlikely to form planets that are comparable to even the low-mass end of observed planet masses. The traps we include are therefore the main traps across the body of the disk that are most likely to play a key role in forming the observed classes of planets.

We determine the ice line's location using an equilibrium chemistry solver, CHEMAPP (distributed by GTT Technologies; http://www.gtt-technologies.de/newsletter), over the range of temperatures and pressures encountered across the disk midplane throughout its evolution. The chemistry calculations are done assuming Solar elemental abundances, and considering the range of metallicities used in our population synthesis calculations. The ice line radius, $r_{il}$ is determined to be the disk radius where the midplane abundances of ice and water vapour are equal. We find the ice line evolves as $r_{il} \sim \dot{M}^{4/9}$, which is the same scaling found in \citet{HP11} who tracked the location in the disk with a midplane temperature of 170 K to determine the ice line's location.

The heat transition, $r_t$, separates the inner region of the disk heated via generalized viscous heating and the outer region of the disk heated through radiation, as discussed in section 2.1. Its location is determined directly in the \citet{Chambers2009} model by equating the midplane temperatures arising from both heating mechanisms. Figure \ref{DiskPlot} shows that both the disk surface density and temperature profile power laws change at the heat transition.

The dead zone outer edge $r_{dz}$, separates an inner laminar region from an outer, turbulent region of the disk \citep{Gammie1996}. MRI-driven turbulence requires a low-level of disk ionization, and within the disk dead zone the surface density is sufficiently high such that ionizing photons are attenuated prior to reaching the disk midplane. \citet{HP10} showed that the outer edge of the dead zone can trap planets due to a sharp increase in dust scale height, with resulting thermal radiation producing an abrupt temperature change leading to planet trapping. 

We refer the reader to \citet{Alessi2017} and \citet{Alessi2018} for detailed descriptions on calculating $r_{dz}$, which closely follows the model presented in \citet{MP2003}. In summary, to determine if MRI-turbulence can be generated at a particular location in the disk, one can equate the MRI growth and damping timescales. This results in a condition for the MRI to be inactive, written in terms of magnetic Elsasser number \citep{Blaes1994, Simon2013},
\begin{equation} \Lambda_0 = \frac{V_A^2}{\eta \Omega_K} \lesssim 1\; ,\end{equation}
where the Alfv\'en speed is $V_A \simeq \alpha_{\rm{turb}}c_s$, and $\eta$ is the magnetic diffusivity, which depends on the electron fraction $x_e$ as follows,
\begin{equation} \eta = \frac{234}{x_e}T^{1/2}\,\rm{cm}^2\,\rm{s}^{-1}\;.\end{equation}
This can be re-written to obtain a critical electron fraction along the midplane, separating the MRI-active and inactive regions ($r_{dz}$). 

The remainder of the calculation of the dead zone's location involves balancing sources and sinks of ionization to determine the electron fraction throughout the disk. We consider ionizing X-rays generated through magnetospheric accretion as the source of ionization in our calculation. This is an update from \citet{Alessi2018}, where in addition to X-rays, interstellar cosmic rays were also considered. X-rays are only considered here since our population results in \citet{Alessi2018} were more consistent with the data when considering X-ray ionization. X-rays being a dominant ionizing source in disks is also supported in astrochemistry calculations that show X-ray ionized models to better reproduce observations \citep{Cleeves2015}, as well as by cosmic ray scattering produced by accretion-generated star and disk winds that can prevent cosmic rays from reaching the disk \citep{Matt2005, Cleeves2013, Frank2014}.

\begin{figure}
\includegraphics[width = 0.45\textwidth]{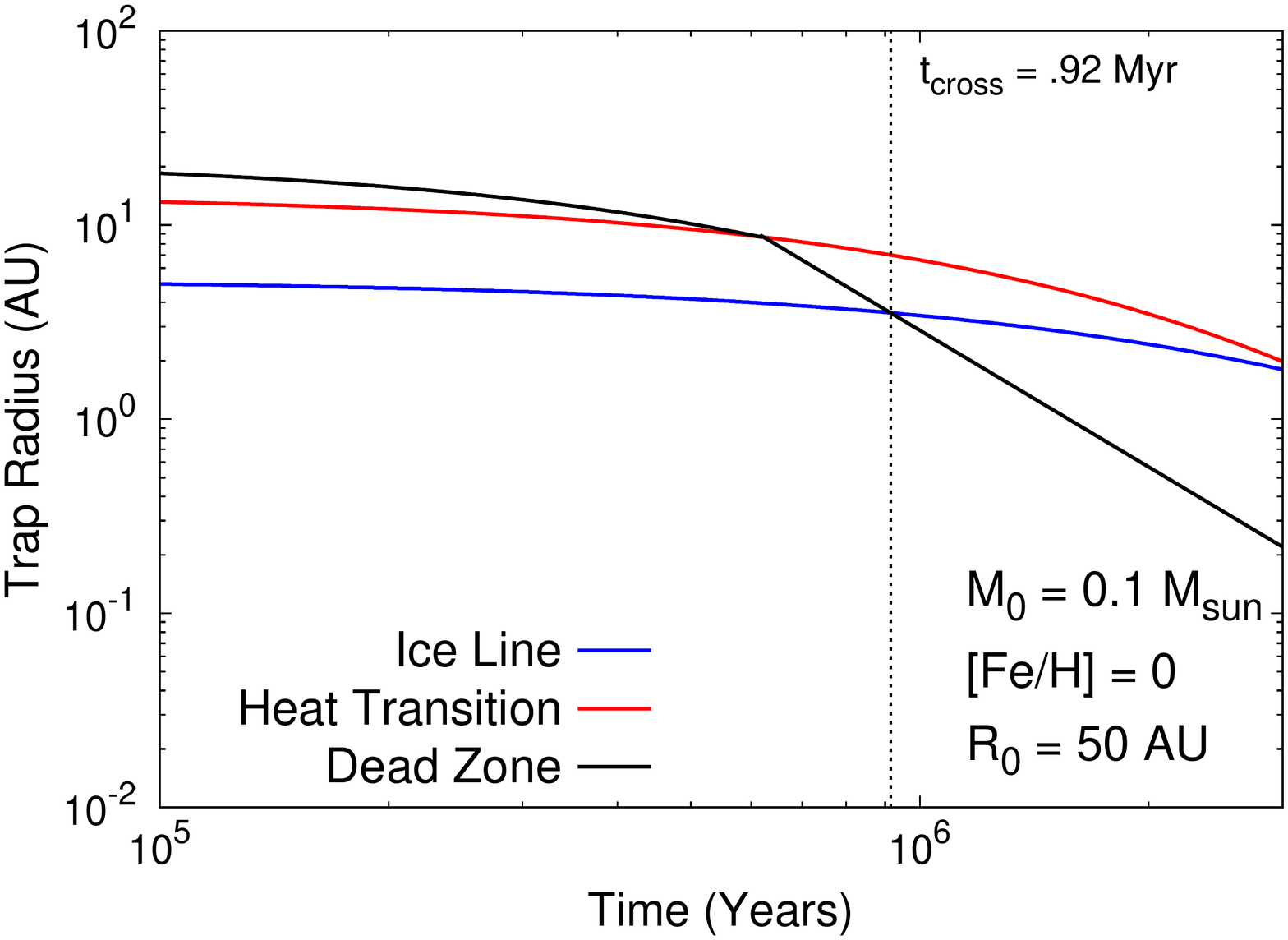}
\caption{The evolution of planet trap radii is shown for a fiducial disk model. The dead zone is initially the outer-most trap in our model, but migrates within the ice line early in the disk's evolution at $t_{\rm{cross}} = 0.92$ Myr labelled in the figure. The heat transition is location outside of the ice line for the entirety of the disk's evolution. Both the the heat transition and ice line traps converge to $\sim$ 1 AU at the disk's 3 Myr lifetime.}
\label{Traps}
\end{figure}

We note that the heat transition is the only trap in our model that has a corresponding transition accounted for in the disk model. We rather use the \citet{Chambers2009} disk model to determine where in the disk the ice line and dead zone are, but there are no transitions in the overall disk surface density or temperature profiles. There is, however, a transition in the surface density of solids, $\Sigma_d$ at the ice line due to the change in fragmentation velocity from the dust evolution model. Including changes in disk surface density and temperature at the ice line and dead zone are not necessary in our calculation as we are not directly computing the type-I migration torques, but rather assume the planets are trapped at the features for the entirety of the type-I migration phase.

In figure \ref{Traps}, we plot the evolution of the three traps in our model for a fiducial disk setup. The ice line and heat transition both converge to $\sim$ 1 AU at the end of the disk's 3 Myr-lifetime, with the heat transition lying outside of the ice line for the entirety of the disk's evolution. The dead zone radius is initially $\sim$ 20 AU, but the trap quickly evolves to the inner disk, crossing the ice line within 1 Myr, and continuing to $\sim 0.02$ AU at the disk's lifetime. The dead zone's evolution is the most drastic of the three because of the nearly horizontal path the magnetospheric accretion-generated X-rays take through the disk, and the resulting attenuation rates being very sensitive to the disk's surface density. Recalling the large change in solid surface density at the ice line and the outer disk depleted of solids by radial drift, the locations of the traps and their evolution will greatly affect the solid accretion stage of planet formation in each of the three traps.

Type-II migration applies to planets that are sufficiently massive to open a gap in the disk structure through gravitational torques. An annular gap is opened if the planet's gravitational torque on disk material exceeds the disk's viscous torque, or if the planet's hill sphere exceeds the disk's pressure scale height \citep{MP2006},
\begin{equation} M_{\rm{GAP}} = M_* \rm{min}\left[3h^3(r_p), \sqrt{40 \alpha h^5(r_p)}\right]\,.\label{GapOpening} \end{equation}
Here, $r_p$ denotes the planet's radius and $h = H/r$ is the disk aspect ratio. 

Planets undergo type-II migration once exceeding the local gap-opening mass, proceeding at a rate determined by the disk's viscous-timescale,
\begin{equation} v_{\rm{mig,II}} = -\nu/r_p\,. \end{equation}
When the planet greatly exceeds its gap-opening mass, achieving masses comparable to the remaining disk mass interior to its orbit ($M_p > M_{\rm{crit}} = \pi r_p^2 \Sigma$), the planet's inertia will slow its migration beneath the disk's viscous rate \citep*{Ivanov1999, HP12},
\begin{equation} v_{\rm{mig,II, slow}} = -\frac{\nu}{r_p\left(1 + M_p/M_{\rm{crit}}\right)}\,.\end{equation}
This migration phase applies to planets that undergo runaway gas accretion. During the type-II migration phases, planets migrate away from the trap they were forming in during the trapped type-I migration regime, and planet orbital radii are not solely determined by the location of the traps.

We initialize our planet formation runs with a 0.01 M$_\oplus$ core situated at the orbital radius of the trap it is forming within. Here we are assuming that type-I migration will quickly migrate planets into a trap within the disk if a distribution of initial orbital radii were instead used. The first growth stage in the core accretion scenario is solid accretion, whereby we model the core's growth to take place via planetesimal accretion or oligarchic growth. The accretion timescale in this phase is \citep{KokuboIda2002},
\begin{equation} \begin{aligned} \tau_{\rm{c,acc}}  \simeq &1.2\times10^5\;\textrm{yr}\;  \left(\frac{\Sigma_d}{10\;\textrm{g cm}^{-2}}\right)^{-1}
\\ & \times \left(\frac{r}{r_0}\right)^{1/2}\left(\frac{M_p}{M_\oplus}\right)^{1/3}\left(\frac{M_*}{M_\odot}\right)^{-1/6} 
\\ & \times\left[\left(\frac{b}{10}\right)^{-1/5}\left(\frac{\Sigma_g}{2.4\times10^3\;\textrm{g cm}^{-2}}\right)^{-1/5} \right.
\\ & \left. \times \left(\frac{r}{r_0}\right)^{1/20}\left(\frac{m}{10^{18} \;\textrm{g}}\right)\right]^2\;,\;\label{Solid_Accretion} \end{aligned}\end{equation}
where $m\simeq10^{18}$ g is the mass of accreted planetesimals and $b\simeq 10$ is a parameter defining the core's feeding zone. The corresponding accretion rate is $\dot{M}_p = M_p/\tau_{\rm{c,acc}}$.

Including the dust evolution model affects our planet formation model in equation \ref{Solid_Accretion}, as we take the local solid surface density $\Sigma_d$ from the dust-to-gas ratio distribution calculated using the \citet{Birnstiel2012} model. In doing so, we are assuming that the distribution of planetesimals will match the disk's dust distribution. This assumption can be justified as streaming instability models have shown planetesimal assembly from dust takes place on short ($\lesssim 10^3$ year) timescales \citep{Johansen2007}. Gravitational dynamics may yet change the planetesimal distribution, however the oligarchic growth phase is short ($\sim10^5-10^6$ years) and we do not consider these effects here.

The second, slow gas accretion phase of the core accretion model begins when the planetesimal accretion rate, and resulting core heating, becomes insufficient to maintain hydrostatic balance in gas surrounding the forming core. The critical core mass where a forming planet transitions from oligarchic growth to slow gas accretion is \citep{Ikoma2000, IdaLin2008, HP14},
\begin{equation}  M_{\rm{c,crit}}   \simeq f_{\rm{c,crit}} \left(\frac{1}{10^{-6} M_\oplus \;\textrm{yr}^{-1}}\frac{dM_p}{dt}\right)^{1/4} M_\oplus \;. \label{CoreCrit}\end{equation}
We set $f_{\rm{c,crit}} = 1.26$, which results from the best-fit envelope opacity of 0.001 cm$^2$ g$^{-1}$ determined in \citet{Alessi2018}, where we explored the full dependence of equation \ref{CoreCrit} on the envelope opacities of forming planets.

The slow gas accretion timescale proceeds at the Kelvin-Helmholtz rate \citep{Ikoma2000},
\begin{equation} \tau_{KH} \simeq 10^c \, \textrm{yr}\left(\frac{M_p}{M_\oplus}\right)^{-d}\;.\label{Gas_Accretion}\end{equation}
We take $c = 7.7$ and $d=2$, as determined by the best-fit envelope opacity from \citet{Alessi2018}, where fits from \citet{Mordasini2014} were used to link the Kelvin-Helmholtz $c$ and $d$ parameters to envelope opacity. The associated gas accretion rate is $\dot{M} = M_p/\tau_{KH}$.

As gas accretion proceeds, the planetary envelope may become sufficiently massive to lose pressure support, whereby the planet will transition into a runaway growth phase. We only consider the Kelvin-Helmholtz timescale in calculating gas accretion rates, and the runaway growth phase is a consequence of $\tau_{KH}$ decreasing as the planet's mass increases. Other works have instead considered the disk-limited accretion phase when determining the gas accretion rate on massive planets (e.g. \citet{Machida2010, Tanigawa2016, Hasegawa2018, Hasegawa2019}). Both approaches can produce similar results depending on the settings of the planet's envelope opacity (or Kelvin-Helmholtz parameters).

The termination of gas accretion onto massive planets is expected physically linked to gap-opening as the local surface density of gas within the planet's feeding zone decreases. We therefore parameterize the maximum mass of a forming planet, following \citet{HP13}, as,
\begin{equation} M_{\rm{max}} = f_{\rm{max}} M_{\rm{gap}} \,. \label{MaximumMass} \end{equation}
Accretion onto planets whose masses exceed $M_{\rm{max}}$ is artificially truncated. The settings of the parameter $f_{\rm{max}}$ range from 1-500, with low values corresponding to planets whose accretion is terminated shortly after gap-opening. It has been shown, however, that substantial gas accretion can be sustained after gap-opening takes place, corresponding to larger $f_{\rm{max}}$ values \citep{Kley1999, Lubow2006, Morbidelli2014}. We highlight that this parameter is the only intrinsic model parameter that we vary in our population synthesis calculations. It is a necessary inclusion to obtain a range of final planet masses corresponding to the data\footnote{We note that in models that assume a disk-limited final accretion stage to truncate gas accretion, the accretion rate onto the planet is parameterized as a fraction of the disk accretion rate. In such models, this parameter serves the same purpose as our $f_{\rm{max}}$ parameter, ultimately setting the final mass of the planet.}.

\citet{Tanigawa2016} show that disk accretion rate limits a planet's gas supply, providing a physical means of terminating gas accretion. This is further explored in \citet{Lee2019}, who consider both disk accretion rate and local hydrodynamic flows to terminate the accretion onto short-period sub-Saturns. Another physical means of terminating gas accretion is through the interaction of the accreting planet's magnetic field and in-falling disk material, resulting in an accretion cross-section that inversely scales with planet mass \citep{Batygin2018, Cridland2018}. Our method of terminating gas accretion simplifies the late stages of planet formation for the purposes of population synthesis calculations.

%% file: AppendixC.tex
\section{Comparing M-a distributions to constant dust-to-gas ratio models}

\begin{figure*}
\centering
\includegraphics[width = 0.97\textwidth]{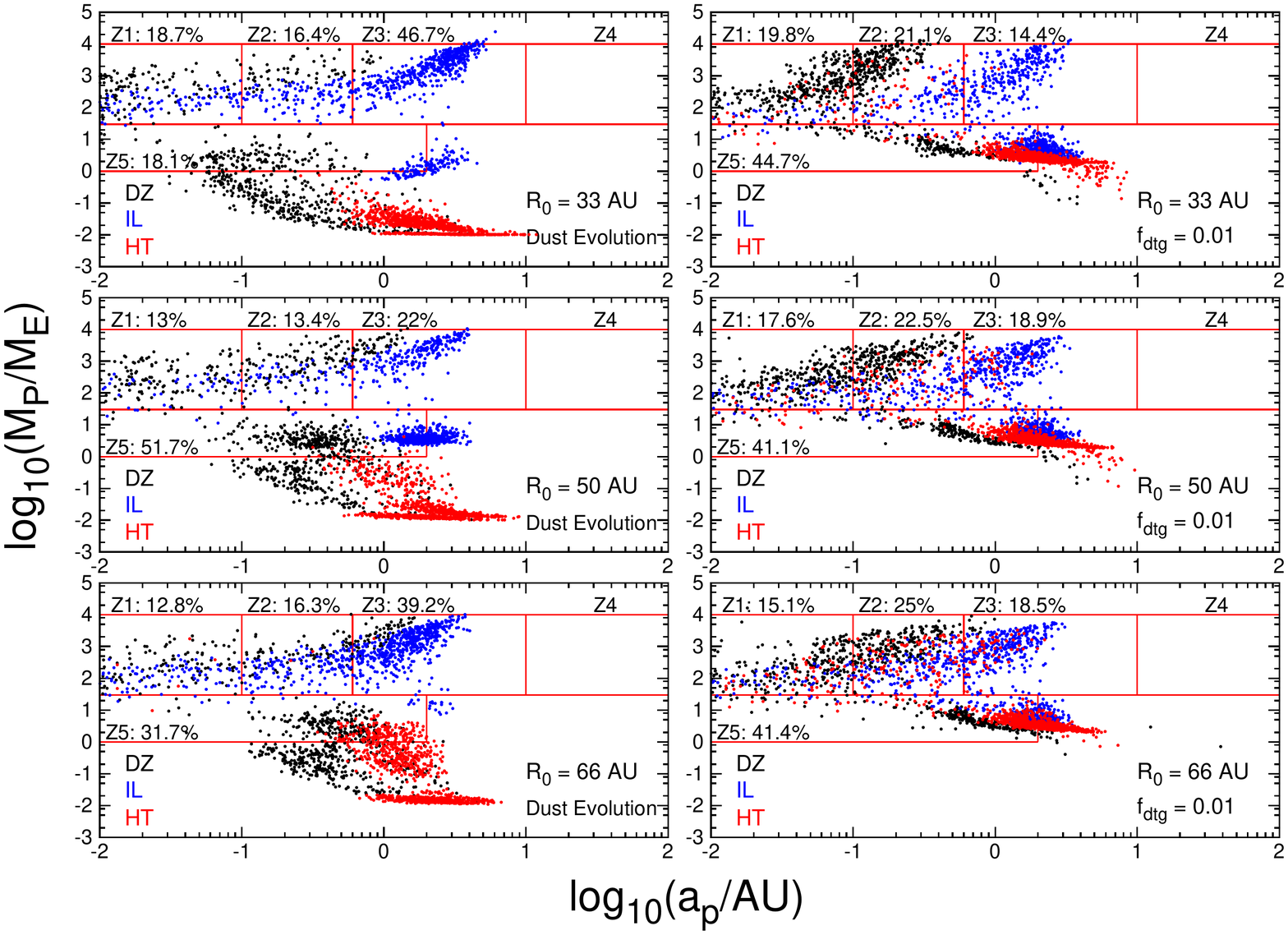}
\caption{\textbf{Left Column:} Planet populations resulting from the full dust evolution model are shown. We consider a range of initial disk radii ($R_0$) between the top (33 AU), middle (the fiducial 50 AU setting), and bottom panels (66 AU). \textbf{Right Column:} Populations resulting from an assumed constant dust-to-gas ratio of 0.01 (neglecting dust evolution as in \citet{Alessi2018}) are shown for comparison, using the same initial disk radii settings.}
\label{Pop_Compare}
\end{figure*}

In figure \ref{Pop_Compare}, we plot M-a distributions of planet populations resulting from two sets of models with different treatments of dust evolution: those with a full dust evolution treatment of \citet{Birnstiel2012} (left column), and those that neglect dust evolution and assume a constant dust-to-gas ratio of 0.01 (right column). We include both models to compare this work's dust treatment to the assumed constant $f_{\rm{dtg}}$ models from \citet{Alessi2018}, and to see the effects the dust evolution model has on our planet populations. For both sets of models, we include a small (33 AU) and large (66 AU) initial disk radius setting, in addition to the fiducial 50 AU models.

The effects of including the dust evolution model are readily seen when compared to the constant dust-to-gas ratio treatment of \citet{Alessi2018}. We first emphasize that the super Earth and warm Jupiter frequencies are insensitive to the initial disk radius setting when a constant dust-to-gas ratio is assumed. Therefore, the disk radius-dependent trade-off between the super Earth and warm Jupiter populations formed at the ice line is only encountered when the full dust evolution treatment is included. 

Inclusion of the dust evolution model also results in more short-period super Earths being formed from the dead zone trap. This is largely a result of the delayed growth in the dead zone, whereby the trap itself needs to migrate to the inner region of the disk before solid accretion can take place. This work's populations that include dust evolution can form super Earths with orbital radii down to $\sim$ 0.03 AU in the case of the smallest disk sizes, and down to $\sim$ 0.08 AU for the fiducial $R_0$ = 50 AU case. The constant dust-to-gas ratios of \citet{Alessi2018} were not able to form short-period, low mass super Earths. This is seen in figure \ref{Pop_Compare} as the constant dust-to-gas ratio models all produce super Earths with orbital periods between $\sim$ 0.8 - 2 AU. Including the dust evolution treatment therefore results in improved population results, as we are readily able to produce shorter-period super Earths, filling out a region of the M-a diagram that is densely populated with observed planets. 

Including dust evolution also results in many sub-Earth mass planets, which were not encountered in the constant dust-to-gas ratio cases. This is due to the solid depletion of the outer disk due to radial drift, and corresponding inefficient solid accretion in a subset of the planet formation runs that does not take place when $f_{\rm{dtg}}$ is held constant. The constant dust-to-gas ratio models additionally have an overall larger gas giant formation frequency because of this.

In the case of a small disk (33 AU), the constant dust-to-gas ratio population shows a separation between the shorter period gas giants formed in the dead zone and the larger period gas giants formed in the ice line - one of the main results of \citet{Alessi2018}. This feature is not seen in any of the models that include dust evolution, nor is it seen in the constant dust-to-gas ratio models where a larger initial disk size is considered.